\documentclass[10pt, twocolumn, twoside]{IEEEtran}

\ifCLASSOPTIONcompsoc
  \usepackage[nocompress]{cite}
\else
  \usepackage{cite}
\fi
\ifCLASSINFOpdf
  \usepackage[pdftex]{graphicx}
  \graphicspath{{../figures/}}
  \DeclareGraphicsExtensions{.pdf,.jpeg,.png}
\else
\fi
\usepackage{amsmath}
\usepackage{amsfonts}
\usepackage{bm}
\usepackage{array}
\usepackage{url}

\newcommand\MYhyperrefoptions{bookmarks=true,bookmarksnumbered=true,
pdfpagemode={UseOutlines},plainpages=false,pdfpagelabels=true,
colorlinks=true,linkcolor={black},citecolor={black},urlcolor={black},
pdftitle={My PDF title},
pdfsubject={Typesetting},
pdfauthor={Savvas Panagiotou},
pdfkeywords={Computer Society, IEEEtran, journal, LaTeX, paper,
             template}}
\ifCLASSINFOpdf
\usepackage[\MYhyperrefoptions,pdftex]{hyperref}
\else
\usepackage[\MYhyperrefoptions,breaklinks=true,dvips]{hyperref}
\usepackage{breakurl} 
\fi

\usepackage{xcolor}

\usepackage[capitalise]{cleveref}
\usepackage{siunitx}
\usepackage{numprint}
\usepackage{multirow}
\usepackage{lipsum}
\usepackage{orcidlink}
\usepackage{tikz}
\usetikzlibrary{positioning, spy}

\begin{document}
%
\title{Denoising Diffusion Post-Processing for Low-Light Image Enhancement}
%
%

\author{Savvas~Panagiotou \orcidlink{0009-0009-6398-8427},~and~Anna~S.~Bosman \orcidlink{0000-0003-3546-1467},~\IEEEmembership{Member,~IEEE}
\IEEEcompsocitemizethanks{\IEEEcompsocthanksitem S. Panagiotou and A.S. Bosman are with the Department of Computer Science, University of Pretoria, Pretoria, Gauteng, South Africa. \protect\\
E-mail:~savva.panagiotou@gmail.com,~anna.bosman@up.ac.za}
}

\IEEEtitleabstractindextext{%
\begin{abstract}

Low-light image enhancement (LLIE) techniques attempt to increase the visibility of images captured in low-light scenarios. However, as a result of enhancement, a variety of image degradations such as noise and color bias are revealed. Furthermore, each particular LLIE approach may introduce a different form of flaw within its enhanced results. To combat these image degradations, post-processing denoisers have widely been used, which often yield oversmoothed results lacking detail. We propose using a diffusion model as a post-processing approach, and we introduce Low-light Post-processing Diffusion Model (LPDM) in order to model the conditional distribution between under-exposed and normally-exposed images. We apply LPDM in a manner which avoids the computationally expensive generative reverse process of typical diffusion models, and post-process images in one pass through LPDM. Extensive experiments demonstrate that our approach outperforms competing post-processing denoisers by increasing the perceptual quality of enhanced low-light images on a variety of challenging low-light datasets. Source code is available at \url{https://github.com/savvaki/LPDM}.

\end{abstract}

}

\maketitle

\IEEEdisplaynontitleabstractindextext

%
\IEEEpeerreviewmaketitle

\ifCLASSOPTIONcompsoc
\IEEEraisesectionheading{\section{Introduction}\label{sec:introduction}}
\else
\section{Introduction}
\label{sec:introduction}
\fi

%
%
%
%
\IEEEPARstart{T}{he} task of low-light image enhancement (LLIE) aims to improve the visibility of images which are captured under low-light conditions. Under-exposed images are often degraded in a variety of ways in addition to their lack of visibility. Notably, low-light regions of an image typically contain degraded color information, a lack of detail as well as intensive noise. LLIE techniques aim to brighten low-light regions of an image while maintaining color accuracy and minimizing noise. The demand for brightening and enhancing low-light images often arises due to many downstream algorithms only being performant on images with high-visibility \cite{ref:LIME}. Some of these downstream tasks include object detection~\cite{ref:low-light-object-detection}, facial recognition~\cite{ref:ve-lol}, surveillance~\cite{ref:llnet-2016} and semantic segmentation~\cite{ref:sgz}.

Simply adjusting the contrast of low-light images using a technique such as histogram equalization \cite{ref:ahe1} is often insufficient due to the amplification of noise \cite{ref:LIME,ref:learning-to-see}. Learning-based methods have emerged which significantly outperform traditional methods. However, even the state-of-the-art deep learning (DL) techniques still introduce a variety of artifacts in different scenarios \cite{ref:kind++}.

Existing denoising techniques can be applied to denoise low-light images either before or after contrast enhancement \cite{ref:deep-ll-post-denoiser, ref:pre-denoising-then-contrast}. These denoising techniques range from low-pass filters and algorithms such as block matching and 3D filtering (BM3D) \cite{ref:bm3d}, to state-of-the-art DL denoisers~\cite{ref:deep-ll-post-denoiser, ref:nafnet-denoiser}. Despite denoisers significantly reducing noise, they often introduce blurriness into the denoised output. As a result, removing the amplified noise in a brightened low-light image often comes at the cost of removing detail, especially in high-frequency regions of the image.

\begin{figure}[t!]
  \setlength\tabcolsep{2pt}
  \tiny
  \centering
  \resizebox{\columnwidth}{!}{
  \begin{tabular}{ccc}
   Input &
   BIMEF \cite{ref:bimef }&
   BIMEF + LPDM (Ours) \\
   \includegraphics[width=1in]{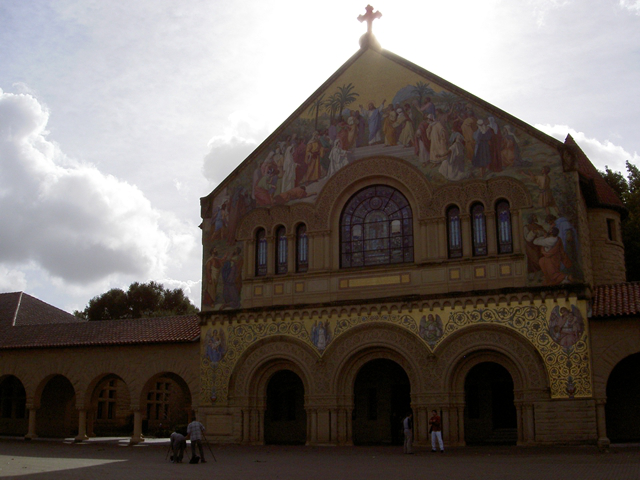}
  &

   \begin{tikzpicture}
    \begin{scope}[
      node distance = 1mm,
          inner sep = 0pt,spy using outlines={rectangle, red, magnification=2,}
                          ]
    \node (n0)  {\includegraphics[width=1in]{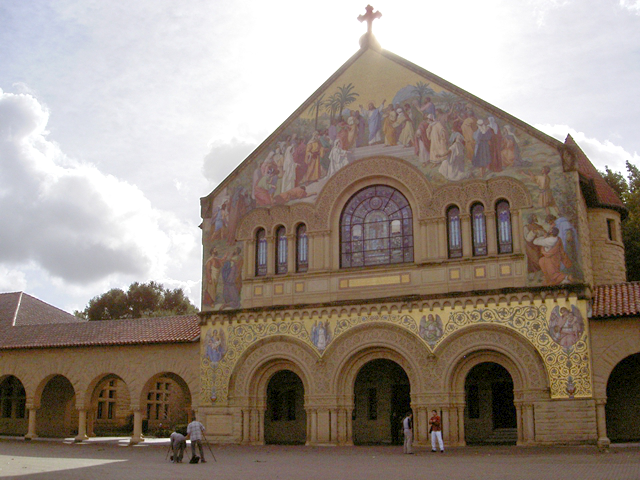}};

    \spy [blue,size=1cm] on (5pt,-15pt) in node[below right=of n0.north west];
    \end{scope}
  \end{tikzpicture}

   &
    
   \begin{tikzpicture}
    \begin{scope}[
      node distance = 1mm,
          inner sep = 0pt,spy using outlines={rectangle, red, magnification=2,}
                          ]
    \node (n0)  {\includegraphics[width=1in]{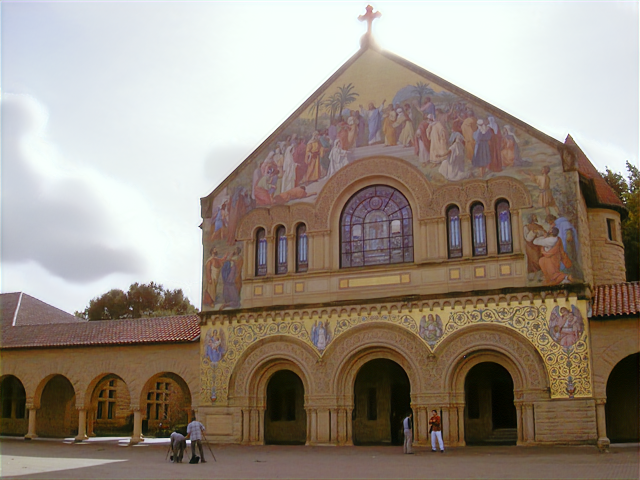}};

    \spy [blue,size=1cm] on (5pt,-15pt) in node[below right=of n0.north west];
    \end{scope}
  \end{tikzpicture}

   \\
   Input&
   LIME \cite{ref:LIME} &
   LIME + LPDM (Ours) \\
   \includegraphics[width=1in]{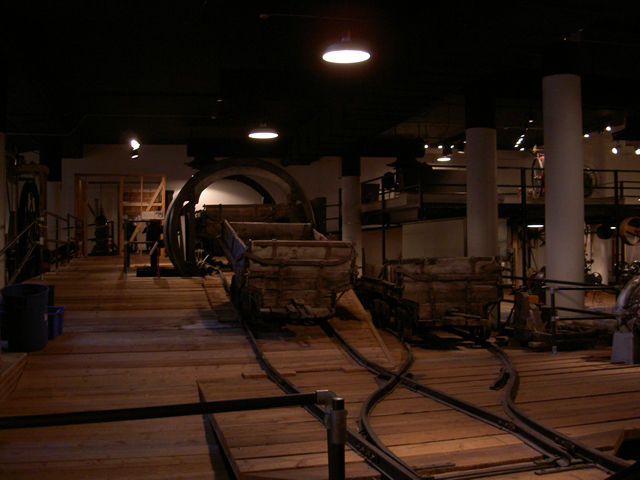} &

   \begin{tikzpicture}
    \begin{scope}[
      node distance = 1mm,
          inner sep = 0pt,spy using outlines={rectangle, red, magnification=2,}
                          ]
    \node (n0)  {\includegraphics[width=1in]{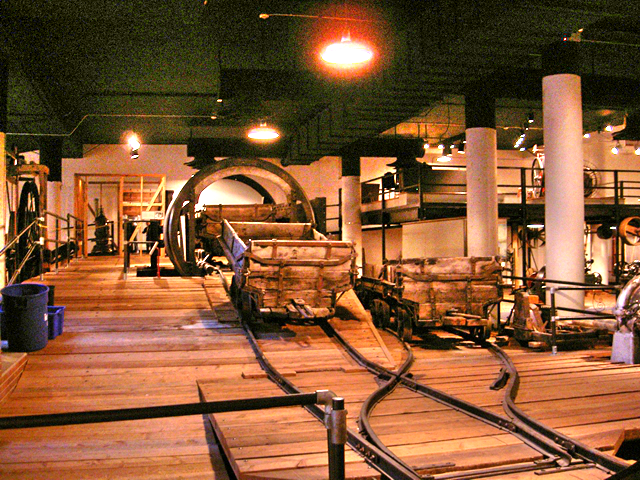}};

    \spy [blue,size=1cm] on (-5pt,15pt) in node[above left=of n0.south east];
    \end{scope}
  \end{tikzpicture}
   
   &
   
   \begin{tikzpicture}
    \begin{scope}[
      node distance = 1mm,
          inner sep = 0pt,spy using outlines={rectangle, red, magnification=2,}
                          ]
    \node (n0)  {\includegraphics[width=1in]{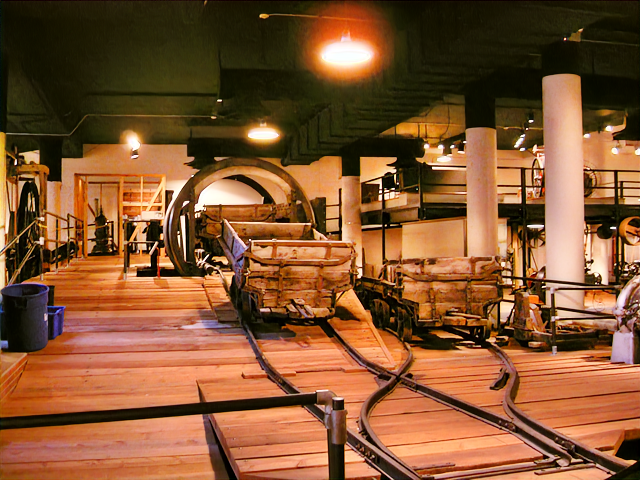}};

    \spy [blue,size=1cm] on (-5pt,15pt) in node[above left=of n0.south east];
    \end{scope}
  \end{tikzpicture}
  \\
   Input &
   LLFlow \cite{ref:llflow}&
   LLFlow + LPDM (Ours) \\
   \includegraphics[width=1in]{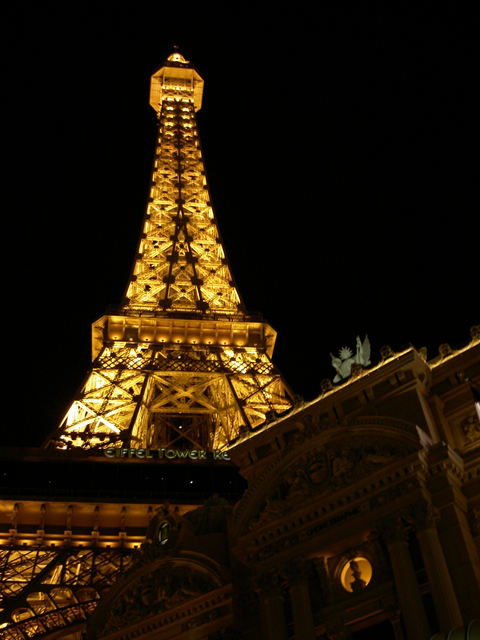} &
   \begin{tikzpicture}
    \begin{scope}[
      node distance = 1mm,
          inner sep = 0pt,spy using outlines={rectangle, red, magnification=2,}
                          ]
    \node (n0)  {\includegraphics[width=1in]{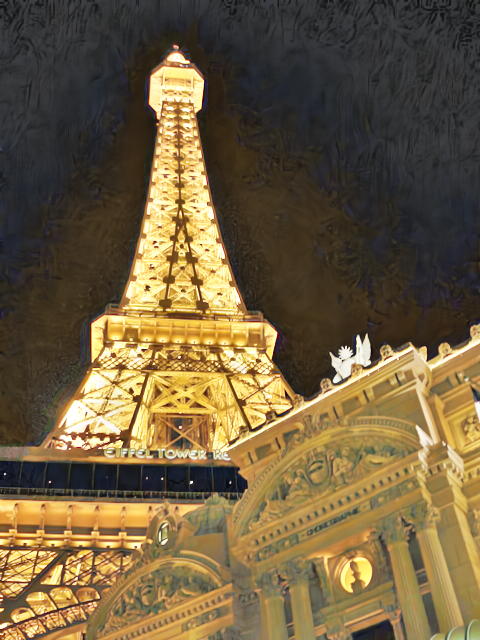}};

    \spy [blue,size=1cm] on (0pt,5pt) in node[below left=of n0.north east];
    \end{scope}
  \end{tikzpicture}
   
    &
   
   \begin{tikzpicture}
    \begin{scope}[
      node distance = 1mm,
          inner sep = 0pt,spy using outlines={rectangle, red, magnification=2,}
                          ]
    \node (n0)  {\includegraphics[width=1in]{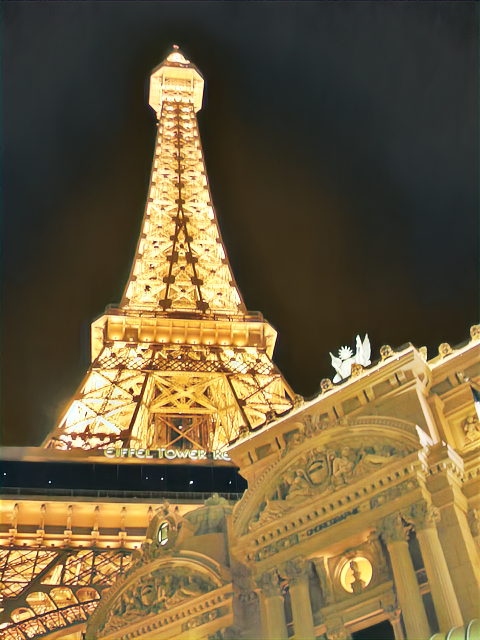} };

    \spy [blue,size=1cm] on (0pt,5pt) in node[below left=of n0.north east];
    \end{scope}
  \end{tikzpicture}
   
   \\
   Input &
   LLFormer \cite{ref:llformer}&
   LLFormer + LPDM (Ours) \\
   \includegraphics[width=1in]{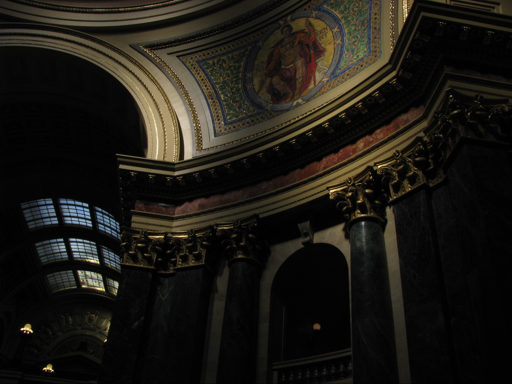} &

   \begin{tikzpicture}
    \begin{scope}[
      node distance = 1mm,
          inner sep = 0pt,spy using outlines={rectangle, red, magnification=2,}
                          ]
    \node (n0)  { \includegraphics[width=1in]{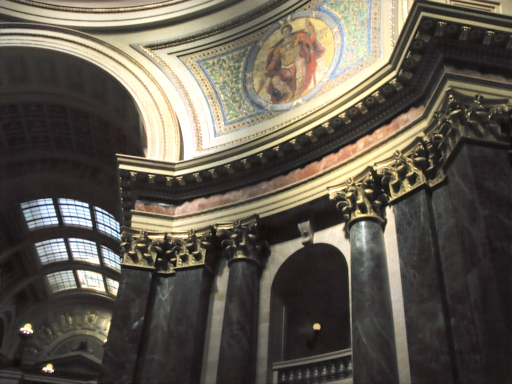} };

    \spy [blue,size=1cm] on (10pt,19pt) in node[above left=of n0.south east];
    \end{scope}
  \end{tikzpicture}
   &
   \begin{tikzpicture}
    \begin{scope}[
      node distance = 1mm,
          inner sep = 0pt,spy using outlines={rectangle, red, magnification=2,}
                          ]
    \node (n0)  { \includegraphics[width=1in]{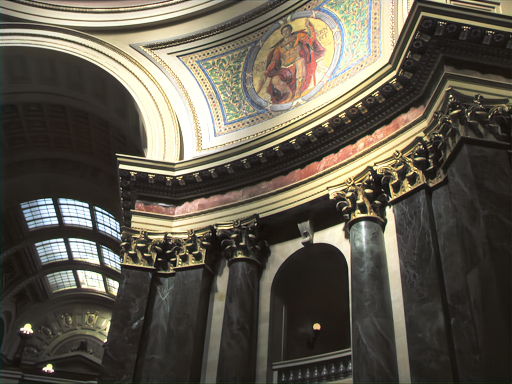} };

    \spy [blue,size=1cm] on (10pt,19pt) in node[above left=of n0.south east];
    \end{scope}
  \end{tikzpicture}
   
   \\
   Input &
   EnlightenGAN \cite{ref:enlightengan} &
   EnlightenGAN + LPDM (Ours) \\
   \includegraphics[width=1in]{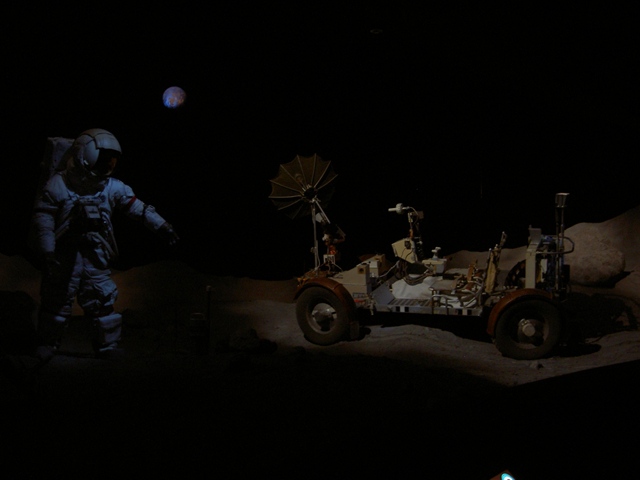} 
   &

   \begin{tikzpicture}
    \begin{scope}[
      node distance = 1mm,
          inner sep = 0pt,spy using outlines={rectangle, red, magnification=2,}
                          ]
    \node (n0)  { \includegraphics[width=1in]{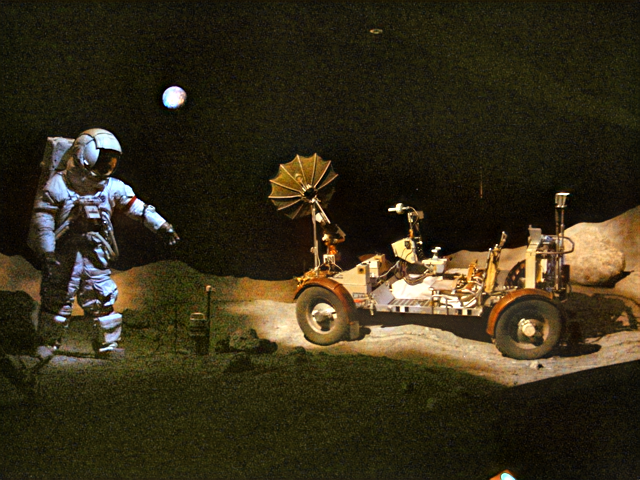} };

    \spy [blue,size=1cm] on (-3pt,7pt) in node[below left=of n0.north east];
    \end{scope}
  \end{tikzpicture}
   
   &
  
  \begin{tikzpicture}
    \begin{scope}[
      node distance = 1mm,
          inner sep = 0pt,spy using outlines={rectangle, red, magnification=2,}
                          ]
    \node (n0)  {\includegraphics[width=1in]{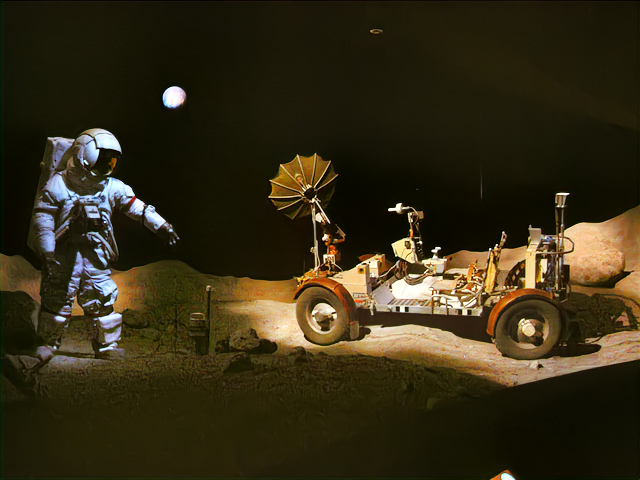}  };

    \spy [blue,size=1cm] on (-3pt,7pt) in node[below left=of n0.north east];
    \end{scope}
  \end{tikzpicture}
   
   \\
   Input &
   KinD++ \cite{ref:kind++} &
   KinD++ + LPDM (Ours) \\
   \includegraphics[width=1in]{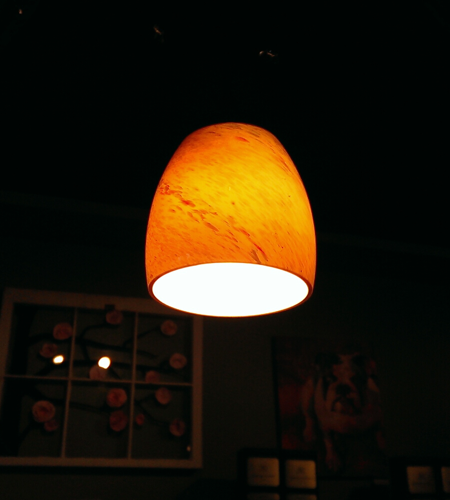} &

   \begin{tikzpicture}
    \begin{scope}[
      node distance = 1mm,
          inner sep = 0pt,spy using outlines={rectangle, red, magnification=2,}
                          ]
    \node (n0)  {  \includegraphics[width=1in]{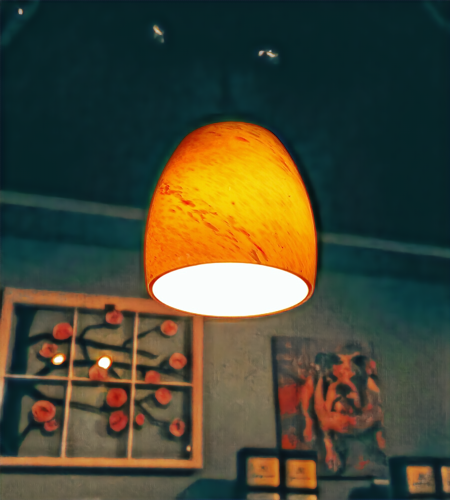}  };

    \spy [blue,size=1cm] on (15pt,-10pt) in node[below right=of n0.north west];
    \end{scope}
  \end{tikzpicture}
   
   &
   
   \begin{tikzpicture}
    \begin{scope}[
      node distance = 1mm,
          inner sep = 0pt,spy using outlines={rectangle, red, magnification=2,}
                          ]
    \node (n0)  { \includegraphics[width=1in]{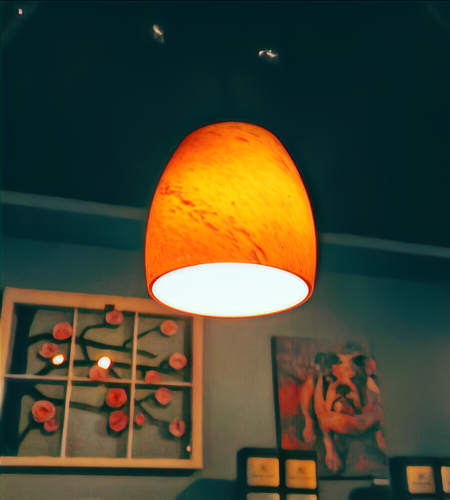}};

    \spy [blue,size=1cm] on (15pt,-10pt) in node[below right=of n0.north west];
    \end{scope}
  \end{tikzpicture}
   \\
   Input &
   URetinex-Net \cite{ref:uretinexnet} &
   URetinex-Net + LPDM (Ours) \\
   \includegraphics[width=1in]{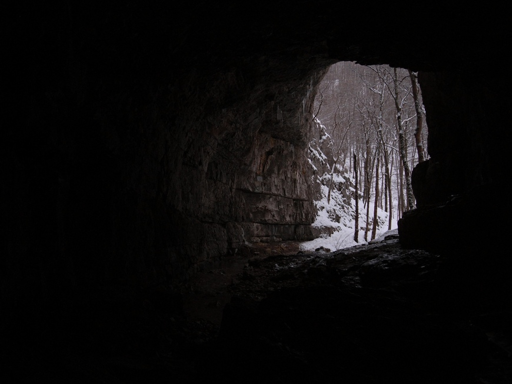} &

    \begin{tikzpicture}
      \begin{scope}[
        node distance = 1mm,
            inner sep = 0pt,spy using outlines={rectangle, red, magnification=2,}
                            ]
      \node (n0)  { \includegraphics[width=1in]{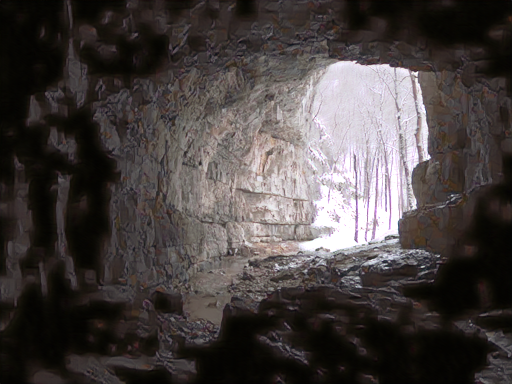} };

      \spy [blue,size=1cm] on (15pt,0pt) in node[above right=of n0.south west];
      \end{scope}
    \end{tikzpicture}
    &
   
   \begin{tikzpicture}
    \begin{scope}[
      node distance = 1mm,
          inner sep = 0pt,spy using outlines={rectangle, red, magnification=2,}
                          ]
    \node (n0)  { \includegraphics[width=1in]{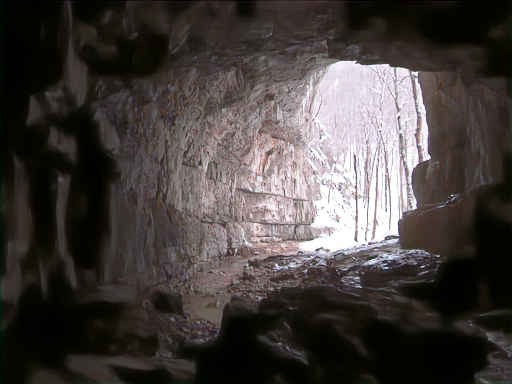}  };

    \spy [blue,size=1cm] on (15pt,0pt) in node[above right=of n0.south west];
    \end{scope}
  \end{tikzpicture}
   
   \\ 
   Input &
   ZeroDCE++ \cite{ref:zero-dce++} &
   ZeroDCE++ + LPDM (Ours) \\
   \includegraphics[width=1in]{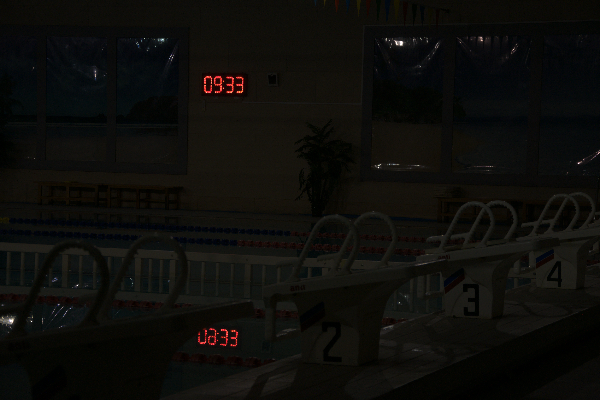}
    &
   
   \begin{tikzpicture}
    \begin{scope}[
      node distance = 1mm,
          inner sep = 0pt,spy using outlines={rectangle, red, magnification=2,}
                          ]
    \node (n0)  { \includegraphics[width=1in]{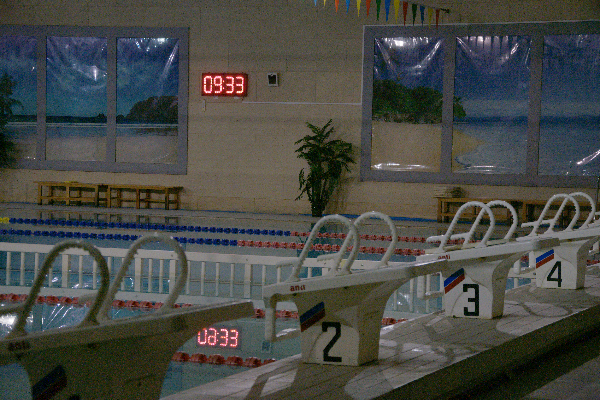}   };

    \spy [blue,size=1cm] on (7pt,5pt) in node[above right=of n0.south west];
    \end{scope}
  \end{tikzpicture}
   &
   \begin{tikzpicture}
    \begin{scope}[
      node distance = 1mm,
          inner sep = 0pt,spy using outlines={rectangle, red, magnification=2,}
                          ]
    \node (n0)  { \includegraphics[width=1in]{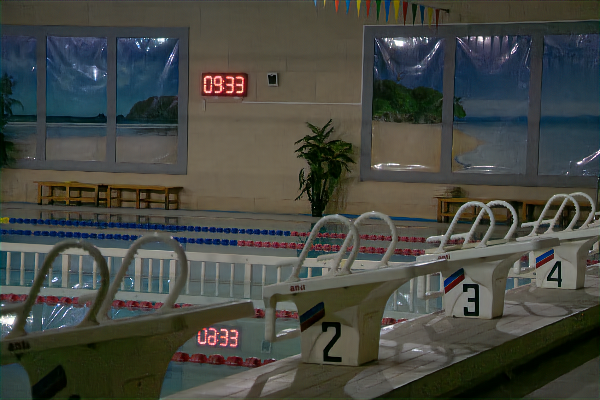}  };

    \spy [blue,size=1cm] on (7pt,5pt) in node[above right=of n0.south west];
    \end{scope}
  \end{tikzpicture}
  \end{tabular}
  }
  
  \caption{Qualitative results of our proposed approach on a variety of evaluation datasets for a variety of LLIE methods. \textbf{Left}: the input test image. \textbf{Middle}: the output of the listed LLIE method for the input image. \textbf{Right}: the result of applying our proposed post-processing LPDM to the image from the middle column. The LPDM parameters used are $\phi = 300$ and $s = 30$.
  \\
  }
  \label{fig:intro}
\end{figure}

We propose a post-processing conditional diffusion model (DM) \cite{ref:diffusion:ddpm} with the capability of removing unwanted noise and other distortions in brightened low-light images. We name our conditional model \textbf{L}ow-light \textbf{P}ost-processing \textbf{D}iffusion \textbf{M}odel (LPDM). The effect of post-processing using LPDM is displayed in \cref{fig:intro}. Our technique is able to avoid the computationally expensive generative diffusion reverse process and denoise a given image in one pass through the model. Furthermore, LPDM is often able to improve the sharpness and color accuracy of the enhanced image. In summary, our contributions are as follows:
\begin{enumerate}
  \item We introduce a method of applying DMs as a post-processing technique in the LLIE pipeline. Our framework is able to circumvent the computationally expensive iterative reverse process of DMs and denoise images in one pass through the model. 
  \item We demonstrate that our DM improves existing state-of-the-art LLIE techniques on popular low-light datasets including challenging unpaired test sets. 
  \item In addition to simple denoising, we demonstrate that our method is able to cope with a variety of different artifacts and color distortions, yielding superior results to existing denoisers for LLIE.
\end{enumerate}

The remainder of this paper is structured as follows: \cref{sec:related-llie} provides background information on LLIE;  \cref{sec:related-diffusion-models} provides background information on DMs; \cref{sec:methodology} outlines preliminary mathematical notation and describes the proposed framework in detail; \cref{sec:experiment} contains the experimental setup and results for this work, including an ablation study; finally, conclusions are drawn in \cref{sec:conclusion}.

\section{Low-Light Image Enhancement}
\label{sec:related-llie}

LLIE techniques have existed for many decades and can be divided into non-learning-based methods and learning-based methods. Popular examples of traditional techniques which do not require learning from data include variants of histogram equalization (HE) \cite{ref:ahe1, ref:ahe2} and gamma correction (GC) \cite{ref:AGC}. HE adjusts the global contrast of an image via a single transformation function. However, low-light images often require contrast enhancements that vary dynamically depending on local regions of the image. Thus, techniques such as GC adjust an image via a non-linear per-pixel transform to brighten dark regions while leaving bright regions relatively unaffected. Despite achieving reasonable results, the abovementioned traditional methods often require post-processing techniques in order to deal with amplified noise after enhancement, and struggle to perform well across diverse scenes. 

Another paradigm of LLIE makes use of Retinex theory, where the assumption is that a color image can be separated into reflectance and illumination \cite{ref:retinex}. Non-learning Retinex-based methods such as Low-light Image Enhancement via
Illumination Map Estimation (LIME) \cite{ref:LIME} provide an effective image enhancement approach; however, post-processing denoising is typically still necessary using algorithms such as BM3D \cite{ref:bm3d} which often blurs high-frequency details. Alternative Retinex-based methods reformulate the traditional Retinex model to incorporate an added noise term in order to cater for noise \cite{ref:retinex-added-noise-term}.

DL methods have recently achieved state-of-the-art LLIE performance. Some complexities of catering for a large variety of realistic low-light scenes are abstracted away by a data-driven approach. Most DL architectures are based on Convolutional Neural Networks (CNNs) and more recently, CNNs have been hybridized with  transformer networks~\cite{ref:attention-is-all}. DL methods either opt for incorporating denoising into a single model, or apply denoising as a post-processing step. S-LLNet \cite{ref:llnet-2016} makes use of a learned denoiser which operates sequentially after contrast enhancement. Retinex-net \cite{ref:lol-dataset} incorporates Retinex theory into a DL model with an  optional illumination-varying BM3D denoiser used as post-processing. Raw sensor data are enhanced and denoised via CNN in \cite{ref:learning-to-see}. A variety of loss functions have also been proposed (in addition to the typical $l_1$ and $l_2$ losses) which further penalize networks based on color, smoothness, brightness and perceptual interpretation \cite{ref:llnet-2016,ref:low-light-ssim-loss,ref:low-light-smoothness-loss, ref:zero-dce, ref:kind, ref:kind++, ref:sgz, ref:drbn, ref:perceptual-loss, ref:sarn}. Unsupervised Generative Adversarial Networks (GANs) have also been proposed for LLIE \cite{ref:enlightengan}. Recently, transformer architectures have gained popularity for LLIE which exploit spatial and channel-wise attention mechanisms \cite{ref:restformer, ref:llformer}.  

Another relevant state-of-the-art approach with respect to this work is the DL model LLFlow \cite{ref:llflow}. The LLFlow framework learns the conditional distribution between low-light and normally-exposed images via the generative paradigm of normalizing flow \cite{ref:normalising-flow}. The LLFlow architecture consists of an encoder as well as an invertible network, trained by minimizing negative log likelihood. The encoder of LLFlow produces an illumination-invariant color map as the prior distribution, upon which the invertible network learns to produce a normally-exposed image. 

LPDM proposed in this study also models the conditional distribution between low-light and normally-exposed images; however, we use the diffusion paradigm to achieve this. Furthermore, we repurpose the function of a DM to be used as a noise detector. Therefore, LPDM provides a subtractable estimation of the noise in an image which can further enhance the image. In contrast to LLFlow, LPDM is used as a post-processing step which can be applied regardless of the enhancing step that precedes LPDM.

\section{Diffusion Models}
\label{sec:related-diffusion-models}

A DM is a form of generative model which has recently been shown to generate high-quality samples, outperforming GANs \cite{ref:diffusion:original2015, ref:diffusion:ddpm, ref:diffusion:improved-ddpm,ref:diffusion:ddpm-beat-gans}. DMs iteratively remove small perturbations of noise, typically starting with a sample from an isotropic Gaussian distribution, until they generate a clean data sample. In this way, the  unconditional diffusion process connects a complex data distribution $q(\bm{x}_0)$ to a simpler, analytically tractable, distribution via a Markov chain consisting of a finite number of timesteps $T$ \cite{ref:diffusion:original2015}. The subscript of a sample indicates a timestep in the Markov chain, with $0$ being a clean sample and $T$ being a sample with the maximum amount of noise added. DMs have been successfully used to model both unconditional and conditional distributions \cite{ref:diffusion:improved-ddpm,ref:classifier-free-guidace}. In spite of the impressive results of DMs, the speed of generating samples has always been a drawback due to their iterative reverse process. Attempts have been made to increase sample speed by making the sampling process non-Markovian, as well as moving DMs to the latent space~\cite{ref:diffusion:ddim,ref:ldm}.

We avoid using DMs for sampling normally-exposed images owing to their expensive generative reverse process. Instead, we exploit the ability of DMs to capture complex conditional data distributions. 
In particular, we use DMs to capture the relationship between under-exposed and normally-exposed images. Other work has shown that DMs may be used as backbone feature extractors which predict features based on noisy inputs \cite{ref:vessel-seg,ref:diffuse-morph}. Furthermore, the task of denoising itself has been shown to assist with seemingly unrelated tasks such as semantic segmentation \cite{ref:denoising-pretraining}. The applications of using DMs in the field of LLIE are relatively unexplored, especially considering that LLIE can be posed as a denoising problem.

\section{Methodology}
\label{sec:methodology}
In this work, we propose a technique where a conditional DM is used to remove noise from images which have undergone LLIE. The remainder of this section is structured as follows: in \cref{sec:method:prelim}, the background information about DMs is outlined; the architecture used for LPDM is described in \cref{sec:method:arch}; finally, in \cref{sec:method:framework}, we provide detail of our proposed framework.

\subsection{Preliminaries}
\label{sec:method:prelim}

 DMs make use of a forward process which adds noise to a sample and a reverse process which removes noise. Our goal is to model the conditional data distribution $\bm{x}_0 \sim q(\bm{x}_0 | \bm{c})$, where $\bm{c}$ is an under-exposed image and $\bm{x}_0$ is a normally-exposed image. The forward diffusion process is defined as follows \cite{ref:diffusion:ddpm}:

\begin{equation}
  \begin{aligned}
    q(\bm{x}_{1:T}|\bm{x}_0) &:= \prod_{t=1}^{T}q(\bm{x}_t|\bm{x}_{t-1}),
    \\
    q(\bm{x}_t | \bm{x}_{t-1}) &:= \mathcal{N}(\bm{x}_t; \sqrt{1- \beta_t}\bm{x}_{t-1}, \beta_t \mathbf{I}),
  \end{aligned}
  \label{eq:diffusion:ddpm:q-def}
\end{equation}
  
\noindent where $\beta_t \in (0,1)$ defines a variance to be used at timestep $t$. As seen by the Markov chain in \cref{eq:diffusion:ddpm:q-def}, obtaining a more noisy sample $\bm{x}_t$ is dependent on the previous less-noisy sample $\bm{x}_{t-1}$. Defining extra notation $\alpha_t := 1 - \beta_t$ and $\bar{\alpha}_t :=  \prod_{s=1}^{t} \alpha_{s}$ allows for \cref{eq:diffusion:ddpm:q-def} to be reformulated to be conditioned on the original clean data sample $\bm{x}_0$ \cite{ref:diffusion:ddpm}:

\begin{equation}
  \begin{aligned}
    q(\bm{x}_t | \bm{x}_0) &:= \mathcal{N}(\bm{x}_t; \sqrt{\bar{\alpha}_t}\bm{x}_0, (1 - \bar{\alpha}_t) \mathbf{I}).
  \end{aligned}
  \label{eq:diffusion:ddpm:q-def-final}
\end{equation}

The efficient sampling of an arbitrary $\bm{x}_t$ at any timestep in the Markov chain is possible given $\bm{x}_0$:

\begin{equation}
  \begin{aligned}
    \bm{x}_t &= \sqrt{\bar{\alpha}_t}\bm{x}_0 + \sqrt{1-\bar{\alpha}_t}\bm{\epsilon}, 
  \end{aligned}
  \label{eq:diffusion:ddpm:q_sample}
\end{equation}

\noindent where $\bm{\epsilon} \sim \mathcal{N}(0,1)$ is a random source. The variance schedule is designed such that $\bm{x}_T \approx \mathcal{N}(0,1)$. Conditional DMs model the reverse process $p_\theta(\bm{x}_{t-1}| \bm{x}_t, \bm{c})$ where $\theta$ indicates that the DM is modelled by a neural network parameterized by $\theta$. The conditional DM attempts to maximize the likelihood $p_\theta(\bm{x}_0 | \bm{c})$. The reverse diffusion process is defined by parameterized Gaussian transitions \cite{ref:diffusion:ddpm}:

\begin{equation}
  \begin{aligned}
    p_\theta(\bm{x}_{0:T}| \bm{c}) &:= p(\bm{x}_T) \prod_{t=1}^{T}p_\theta(\bm{x}_{t-1}|\bm{x}_{t}, \bm{c}),
  \\
  p_\theta(\bm{x}_{t-1} | \bm{x}_t, \bm{c}) &:= \mathcal{N}(\bm{x}_{t-1}; \bm{\mu}_\theta(\bm{x}_t, t, \bm{c}), \bm{\Sigma}_\theta(\bm{x}_t, t, \bm{c})).
  \end{aligned}
\label{eq:diffusion:ddpm:p-def}
\end{equation}

 \noindent In order to avoid learning the variance, let $\bm{\Sigma}_\theta(\bm{x}_t, t, \bm{c}) = \sigma_t^2\mathbf{I}$, where $\sigma_t^2 = \frac{1-\bar{\alpha}_{t-1}}{1-\bar{\alpha}_t}\beta_t$ is a time-dependent constant \cite{ref:diffusion:ddpm}. Therefore, the only learnable component is $\bm{\mu}_\theta$. Instead of directly predicting $\bm{\mu}_\theta$, the DM is parameterized in terms of a denoising autoencoder $\bm{\epsilon}_\theta(\bm{x}_t, t, \bm{c})$ where $t = 1, ..., T$. The number of timesteps $T$ is set to a large number (such as $T=1000$) in order for the reverse process to better-approximate a Gaussian distribution \cite{ref:diffusion:ddim}. The corresponding simplified objective is as follows \cite{ref:diffusion:ddpm}:

\begin{equation}
  \begin{aligned}
    L_{DM} = \mathbb{E}_{\bm{x}_0, \bm{c}, \bm{\epsilon}, t}[\lVert \bm{\epsilon} - \bm{\epsilon}_\theta(\sqrt{\bar{\alpha}_t}\bm{x}_0 + \sqrt{1 - \bar{\alpha}_t}\bm{\epsilon}), t, \bm{c} \rVert^2],
  \end{aligned}
  \label{eq:diffusion:ddpm:loss-simple}
\end{equation}

\noindent where $t$ is uniformly sampled from $\{1, ... , T\}$ and $\bm{\epsilon} \sim \mathcal{N}(0,1)$. In simplified terms, $L_{DM}$ guides the DM to predict the underlying $\bm{\epsilon}$ that was involved in sampling $\bm{x}_t$. Given $\bm{x}_t$ and a prediction for $\bm{\epsilon}$ using $\bm{\epsilon}_\theta$, we can calculate an estimate of $\bm{x}_0$ \cite{ref:diffusion:ddpm}:

\begin{equation}
  \begin{aligned}
    \bm{x}_0 \approx \bm{\hat{x}}_0 = \frac{1}{\sqrt{\bar{\alpha}_t}}  \bm{x}_t - \left(\sqrt{\frac{1}{\bar{\alpha}_t} -1}\right)\bm{\epsilon}_\theta(\bm{x}_t, t, \bm{c}).
  \end{aligned}
  \label{eq:diffusion:predict-x0}
\end{equation}
\noindent Further information about the DM sampling process is omitted since it is not used in this work.

\subsection{Diffusion Model Architecture}
\label{sec:method:arch}
The DM architecture $\bm{\epsilon}_\theta$ used for modelling the diffusion process is typically a form of modified U-Net \cite{ref:unet,ref:diffusion:ddpm}. By definition, a U-Net consists of an encoder and a decoder. The encoder contains a set of residual blocks \cite{ref:resnet} followed by downsampling operations which are repeated multiple times until the desired latent resolution is achieved. The decoder consists of residual blocks followed by upsampling operations in order to return the latent encoding to the original input resolution. Each downsampling stage halves the input resolution and each upsampling stage doubles the input resolution. Before each set of residual blocks in the decoder, output from the encoder with the corresponding resolution is concatenated such that spatial context is not lost as a result of the downsampling operations. Between the encoder and decoder is a middle block which contains a specified number of residual blocks in order to process the latent encoding. 

\begin{figure}[t]
  \includegraphics[width=\linewidth]{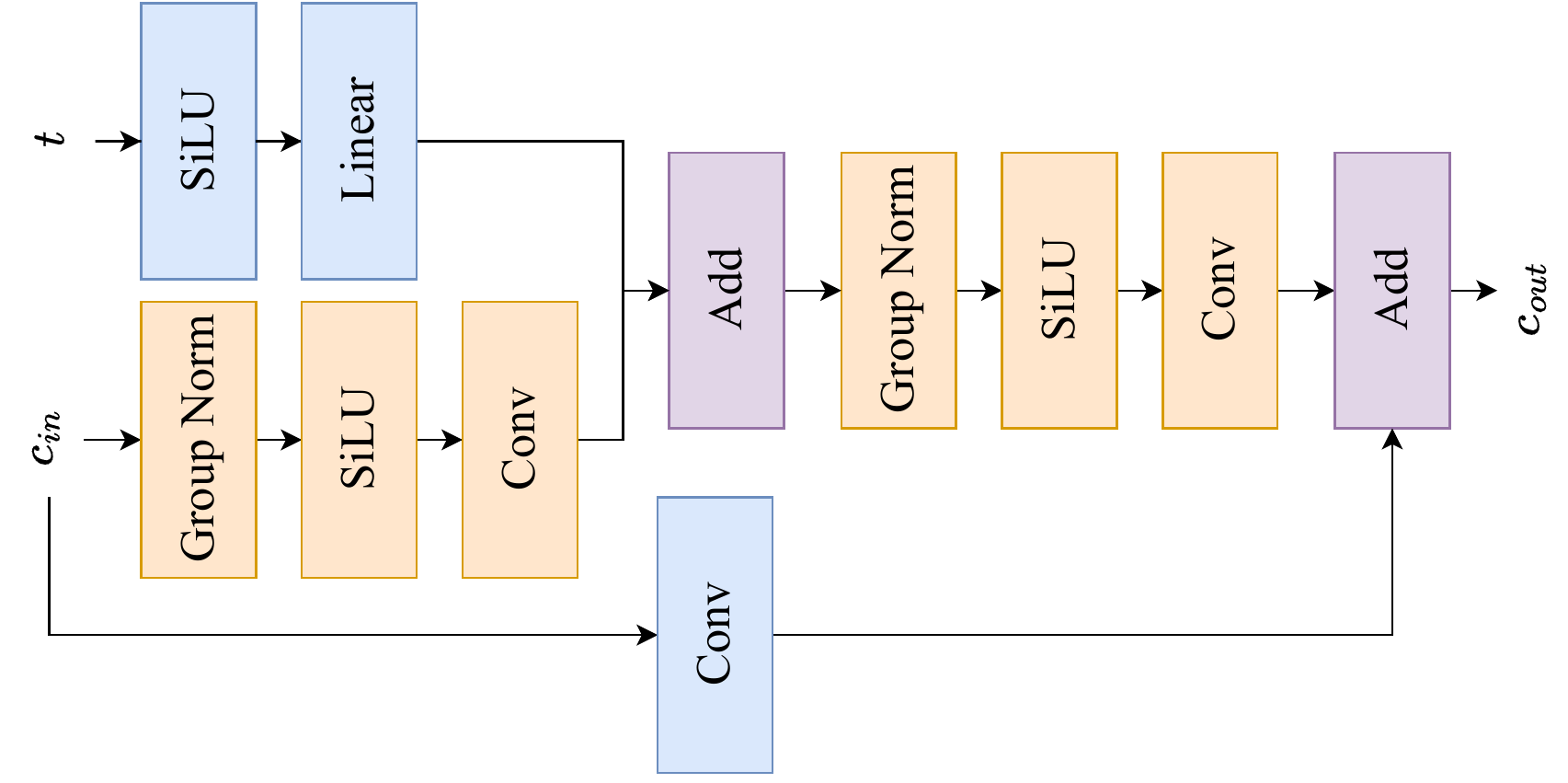}
  \centering
  \caption{Residual block used throughout the DM architecture consisting of a combination of group normalization layers, SiLU activations, convolution layers and addition operations. Both $c_{in}$ and $t$ are inputs to the residual block and represent the channel and timestep-embedded input respectively. Note that $t$ is already in embedded form when it enters the residual block. The output of the residual block is represented by $c_{out}$.}
  \label{fig:res-block}
\end{figure}

The DM operates across all timesteps in the Markov chain using the same parameters. This is made possible by modifying the original U-Net to condition on timestep information represented by sinusoidal positional embeddings as seen in transformer networks \cite{ref:attention-is-all,ref:diffusion:ddpm}. Other notable modifications from the original U-Net are the use of attention mechanisms at different spatial resolutions as well as the use of group normalization \cite{ref:group-norm} within the residual blocks. As seen in \cref{fig:res-block}, the residual blocks of the DM are composed of a combination of group normalization layers, SiLU activations \cite{ref:silu}, convolutional layers and addition operators. A variety of attention mechanisms may be applied at different spatial resolutions, such as global attention \cite{ref:global-attention}, cross-attention \cite{ref:cross-attention} or scaled dot-product attention \cite{ref:attention-is-all}.

\subsection{Proposed Framework}
\label{sec:method:framework}

\begin{figure}
  \setlength\tabcolsep{2pt}
  \small
  \centering
  \begin{tabular}{ccc}
   $\bm{x}_0$ &
   $\bm{c}$ &
   $\bm{\hat{x}}_0^\eta$ \\
   \includegraphics[width=1in]{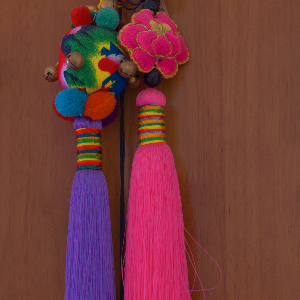} &
   \includegraphics[width=1in]{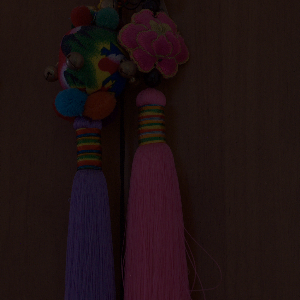} &
   \includegraphics[width=1in]{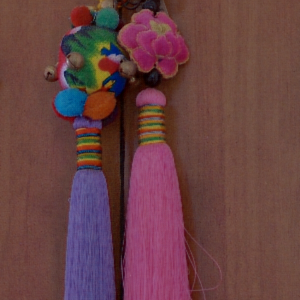} \\
   $\bm{x}_1$ &
   $\bm{x}_{300}$ &
   $\bm{x}_T$ \\
   \includegraphics[width=1in]{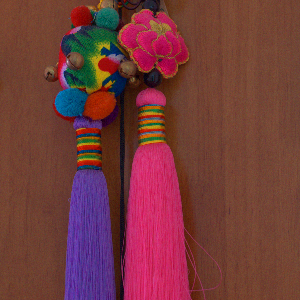} &
   \includegraphics[width=1in]{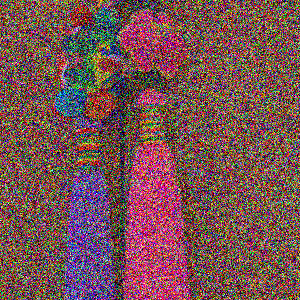} &
   \includegraphics[width=1in]{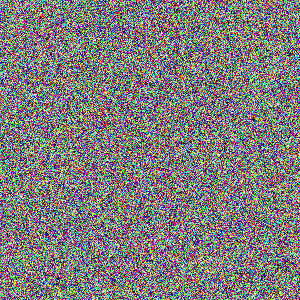} \\
   $\bm{n}_1$ &
   $\bm{n}_{300}$ &
   $\bm{n}_T$ \\
   \includegraphics[width=1in]{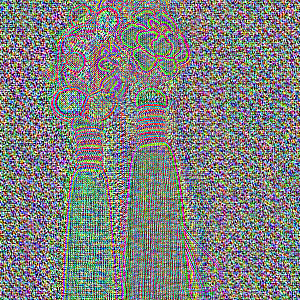} &
   \includegraphics[width=1in]{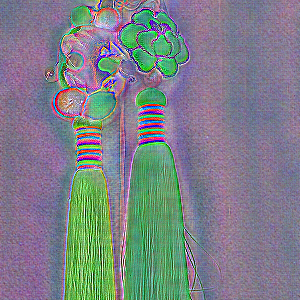} &
   \includegraphics[width=1in]{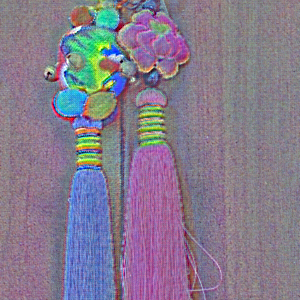} \\
   $\xi(\bm{\hat{x}}_0^\eta, \bm{n}_{300}, 100)$ &
   $\xi(\bm{\hat{x}}_0^\eta, \bm{n}_{300}, 300)$ &
   $\xi(\bm{\hat{x}}_0^\eta, \bm{n}_{300}, T)$ \\
   \includegraphics[width=1in]{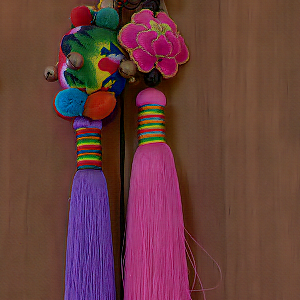} &
   \includegraphics[width=1in]{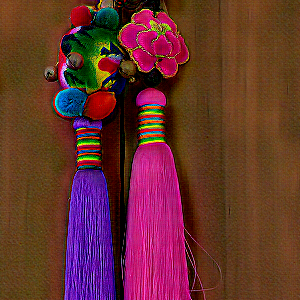} &
   \includegraphics[width=1in]{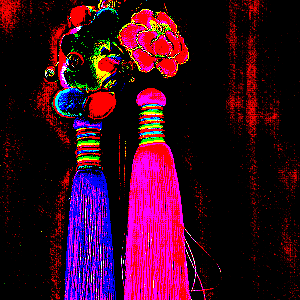} 
  \end{tabular}
  \caption{ Visualization of different components within the diffusion process and the LPDM pipeline. The first row displays a normally-exposed image, under-exposed image and image which has undergone LLIE. The second row demonstrates how noise is added to $\bm{x}_0$ using a linear variance schedule during the training process, with $T = 1000$. The third row demonstrates the effect of different values of $\phi$ in \cref{eq:proposed:noise-estimate}. The fourth row demonstrates the effect of applying \cref{eq:diffusion:predict-x0-mod} with different values of $s$ in order to enhance $\bm{\hat{x}}_0^\eta$.
  }
  \label{fig:phi}
\end{figure}

\begin{figure*}
  \includegraphics[width=\textwidth]{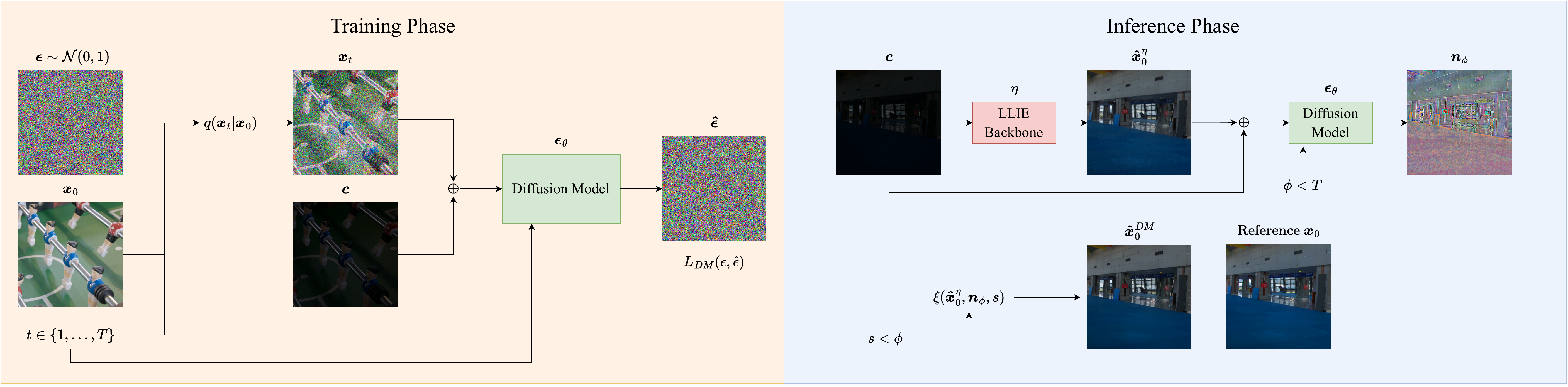}
  \centering
  \caption{Diagram presenting the training phase and inference phase of the LPDM, displayed on the left and right half of the diagram respectively.}
  \label{fig:train-infer}
\end{figure*}

In order to address possible degradations which occur after LLIE, we propose LPDM represented by $\bm{\epsilon}_\theta$. The LPDM model is given $(\bm{x}_t \oplus \bm{c}, t)$ as input and tasked with predicting $\bm{\epsilon}$, where $\bm{x}_t \sim q(\bm{x}_t | \bm{x}_0)$ is a normally-exposed image with noise $\bm{\epsilon} \sim \mathcal{N}(0,1)$ added at a timestep $t$, $\bm{c}$ is the corresponding under-exposed image and $\oplus$ is the concatenation operator.

We follow the common DM training process where the LPDM is exposed to batches of $(\bm{x}_t \oplus \bm{c}, t)$ with randomly sampled $t$ for each sample in a batch. A visualization of the training process is depicted in the left half of \cref{fig:train-infer}. After the LPDM has been trained, the LPDM is applied in a novel manner. Specifically, the LPDM acts as a noise detector given an enhanced low-light image. Let $\eta$ be any low-light image enhancer and let $\bm{\hat{x}}_0^\eta$ be an enhanced image such that  $\bm{\hat{x}}_0^\eta = \eta(\bm{c})$. The LPDM is used to obtain an estimate of the noise present in $\bm{\hat{x}}_0^\eta$:

\begin{equation}
  \begin{aligned}
    \bm{n}_{\phi} &= \bm{\epsilon}_\theta(\bm{\hat{x}}_0^\eta, \phi, \bm{c}),
  \end{aligned}
  \label{eq:proposed:noise-estimate}
\end{equation}

\noindent where $\phi < T$ is a timestep at which we wish to detect noise in $\bm{\hat{x}}_0^\eta$. An important property of \cref{eq:proposed:noise-estimate} is that noise is not added to $\bm{\hat{x}}_0^\eta$; rather, the model is tasked with finding the noise present in $\bm{\hat{x}}_0^\eta$ as a result of enhancement, based on the conditioning $\bm{c}$ and timestep $\phi$. Additionally, the value of $\phi$ is selected at a level such that the underlying structure of the image would be preserved if noise were hypothetically to be added. Thus, a suitable value for $\phi$ is related to the selected variance schedule. \cref{fig:phi} provides visual examples of applying our proposed approach for an example set of $\bm{x}_0$, $\bm{c}$ and $\bm{\hat{x}}_0^\eta$. Row 2 of \cref{fig:phi} demonstrates the effect of selecting different values of $t$ in \cref{eq:diffusion:ddpm:q-def-final} in order to sample a noisy image during the training process. The noise schedule closer to 0 corresponds to less added noise in the image. Therefore, it is reasonable to detect noise in $\bm{\hat{x}}_0^\eta$ at lower levels of~$\phi$ since we do not expect $\bm{\hat{x}}_0^\eta$ to be pure noise. 

Row 3 of \cref{fig:phi} demonstrates the effect of different values of $\phi$. For values of $\phi$ that are too low, the model overestimates the noise present in $\bm{\hat{x}}_0^\eta$. For large values of $\phi$, the model becomes similar to an autoencoder and attempts to predict the input. The reason for the behavior of different levels of $\phi$ can be explained by how the LPDM is trained. For values of $\phi$ close to $T$, the LPDM expects the input to be purely noise, and thus a suitable prediction for $\bm{\epsilon}$ would simply be to predict the input. For low values of $\phi$, the model expects subtle noise in the input image and thus may overdetect noise. We find a good balance to be values of $\phi$ where the background structure of the image is not completely destroyed such as $\phi = 300$ which corresponds to $\bm{x}_{300}$ in row 2 of \cref{fig:phi}.

Once we obtain the estimation of the noise $\bm{n}_{\phi}$, we subtract the noise from $\bm{\hat{x}}_0^\eta$ using a modification of \cref{eq:diffusion:predict-x0}:

\begin{equation}
  \begin{aligned}
    \bm{\hat{x}}_0^{DM} = \xi(\bm{\hat{x}}_0^\eta, \bm{n}_{\phi}, s) := \frac{1}{\sqrt{\bar{\alpha}_s}}  \bm{\hat{x}}_0^\eta - \left(\sqrt{\frac{1}{\bar{\alpha}_s} -1}\right)\bm{n}_{\phi},
  \end{aligned}
  \label{eq:diffusion:predict-x0-mod}
\end{equation}

\noindent where $s$ is a timestep which selects the coefficients according to the variance schedule, and thus balances the degree to which noise is subtracted from $\bm{\hat{x}}_0^\eta$. The final result after LPDM post-processing is represented by $\bm{\hat{x}}_0^{DM}$. Notably, we find that $s$ should be significantly less than $\phi$ in order to subtract the correct amount of noise. As $s \rightarrow \phi$, more of the noise $\bm{n}_{\phi}$ is subtracted from $\bm{\hat{x}}_0^{\eta}$, leading to overcorrections and perhaps further degrading the result. As seen in row 4 of \cref{fig:phi}, the value of $s$ impacts how much correction should be applied. Our technique is able to reduce noise, correct color and improve sharpness as seen when comparing $\bm{\hat{x}}_0^{\eta}$ to $\xi(\bm{\hat{x}}_0^\eta, \bm{n}_{300}, 100)$ in \cref{fig:phi}.

Denoising techniques may be classified as being either blind or non-blind. Blind denoisers do not require the user to specify the level of noise in the input image, whereas non-blind denoisers require the user to specify the noise level. The popular BM3D algorithm is a non-blind approach. Similarly, our approach requires a selection of $s$ to determine to what extent noise should be subtracted, however we find low values of $s$ to be applicable to a wide variety of scenarios.

In summary, our approach requires the specification of a parameter $s$ during application, where $\phi$ may be fixed empirically. We find $\phi = 300$ to be a reasonable choice, and we use this value for all experiments. As $s \rightarrow 0$, the amount of correction lessens. The right half of \cref{fig:train-infer} summarizes the inference process described above.

\section{Experiments}
\label{sec:experiment}
The following subsections outline the experimental setup: 
\cref{sec:experiment:eval-ds} describes the datasets used in this study; \cref{sec:experiment:implementation} defines the configuration of LPDM and the training parameters used for all experiments; \cref{sec:experiment:benchmark-study} provides detail on the LLIE models selected for comparison with LPDM; in order to achieve a fair comparison, we compare our approach to alternative denoising methods described in \cref{sec:experiment:alt-denoisers}; the interpretation of all the results is presented in \cref{sec:experiment:interp-results}; finally, an ablation study is conducted in \cref{sec:experiment:ablation}. 

\subsection{Evaluation Datasets}
\label{sec:experiment:eval-ds}

Paired low-light datasets are challenging to collect due to the requirement of having the scene remain unchanged while camera ISO is adjusted \cite{ref:lol-dataset}. Accordingly, many methods resort to augmenting datasets with synthetic data. Synthetic datasets are typically generated by adjusting the gamma of normally-exposed images and adding simulated noise. We avoid training the LPDM on synthetic datasets in order to ensure that we correctly model the conditional distribution between under-exposed and normally-exposed images. We train on the original paired LOL dataset \cite{ref:lol-dataset}, which contains 485 training images and 15 test images. Alternative versions of LOL exist, however these only add synthetic data such as the extended version of LOL \cite{ref:lolv2} and VE-LOL \cite{ref:ve-lol}. We evaluate our model on the widely-adopted real unpaired test sets LIME~\footnote{LIME \cite{ref:LIME} is both a LLIE technique and an unpaired test dataset, both of which are proposed in the same paper.} (10 images)~\cite{ref:LIME}, DICM (64 images)\cite{ref:dicm}, MEF (17 images)~\cite{ref:mef}, NPE (7 images)~\cite{ref:npe}. We specify the number of images explicitly as previous works use varying subsets of the test sets. The full-reference metrics we adopt are structural similarity index measure (SSIM), peak signal-to-noise ratio (PSNR), mean absolute error (MAE) and learned perceptual image patch similarity (LPIPS) \cite{ref:lpips}. For the unpaired test data we adopt the following no-reference metrics: natural image quality evaluator (NIQE) \cite{ref:niqe}, blind/referenceless image spatial quality evaluator (BRISQUE) \cite{ref:brisque} and the smartphone photography attribute and quality (SPAQ) database \cite{ref:spaq}. All metrics are calculated in the RGB color space unless otherwise stated. 

\subsection{Implementation Details}
\label{sec:experiment:implementation}
A linear variance schedule is used in the range [0.00085, 0.012] for the diffusion process. The value of $T$ is fixed to $1000$ for all experiments. The LPDM is trained on the LOL training set for 6000 training steps using the AdamW optimizer \cite{ref:adamw} with a learning rate of \num{1e-6} and with the AdamW parameters $\beta_1 = 0.9$, $\beta_2 = 0.999$ and $\lambda = 0.01$. The loss function is defined in \cref{eq:diffusion:ddpm:loss-simple}. We use the RGB color space for both low-light and normally-exposed images, and images are converted into the range [-1, 1]. We train on 256 $\times$ 256 random crops with random horizontal flipping.  A batch size of 4 is used with an accumulation of gradients for 8 batches in order to simulate a batch size of 32. 

The U-Net of the LPDM consists of 4 downsampling stages (encoder) and 4 upsampling stages (decoder), with 2 residual blocks per stage. Between the encoder and the decoder is a middle block which processes the latent encoding. The middle block contains 2 residual blocks which surround a scaled dot-product attention layer using 8 attention heads.
We avoid using attention mechanisms at higher resolutions than the final latent encoding in order to conserve memory. Residual blocks within the four downsampling stages output 128, 256, 512 or 512 channels, respectively. Therefore, the residual blocks at the highest resolution output 128 channels, and the latent resolution residual blocks output 512 channels. The residual blocks in the middle block both output 512 channels. The decoder is a reflection of the encoder, where downsampling operations are replaced with upsampling operations and the output of the encoder is concatenated at each upsampling stage.

The implementation of the LPDM method is available online at \url{https://github.com/savvaki/LPDM}.

\subsection{Benchmark Study for LPDM}
\label{sec:experiment:benchmark-study}
The following state-of-the-art LLIE approaches are selected for comparison:  LIME~\cite{ref:LIME}, BIMEF~\cite{ref:bimef}, RetinexNet~\cite{ref:lol-dataset}, EnlightenGAN~\cite{ref:enlightengan}, KinD~\cite{ref:kind}, KinD++~\cite{ref:kind++}, ZeroDCE~\cite{ref:zero-dce}, ZeroDCE++~\cite{ref:zero-dce++}, URetinex-Net~\cite{ref:uretinexnet}, LLFlow~\cite{ref:llflow} and LLFormer~\cite{ref:llformer}. BIMEF and LIME are non-learning-based methods and the remaining methods are all learning-based. The LLFlow, LLFormer, URetinex-Net and RetinexNet methods are trained on the LOL dataset only. The KinD and KinD++ models are trained on LOL with additional custom synthetic data added \cite{ref:kind++}. The ZeroDCE and ZeroDCE++ models are trained on multi-exposure image sets from the SICE \cite{ref:sice} dataset. EnlightenGAN is trained with unpaired groups of low-light and normal-exposure images using data from LOL as well as additional datasets.

Several methods such as KinD, KinD++, URetinex-Net and LLFlow scale their model outputs based on an illumination ratio which involves the ground truth label. In order to achieve a fair comparison, we do not use the ground truths to scale these model outputs. Instead, we treat the LOL test set as unpaired data as would be the case in real-life scenarios. Similar to previous approaches, we fix the parameters of the LIME algorithm to $\alpha = 0.15$ and $\sigma = 2$ and $\gamma = 0.8$.

We compare the abovementioned LLIE approaches with and without our proposed LPDM. For the remainder of this work, $\textnormal{LPDM}_s$ represents the LPDM approach applied with parameter $s$, defined in \cref{sec:method:framework}. All LPDM results fix $\phi = 300$, and we report two values of $s$ which are 15 and 30. \cref{tab:lol-results} contains results with and without LPDM on the LOL test set represented as $\eta + \textnormal{LPDM}_s$. In addition to the LOL dataset evaluation, we provide qualitative and quantitative results on the abovementioned unpaired test sets. The unpaired test metrics can be found in \cref{tab:unpaired-results}.

\sisetup{detect-weight,mode=text}
\renewrobustcmd{\bfseries}{\fontseries{b}\selectfont}
\renewrobustcmd{\boldmath}{} 
\newrobustcmd{\B}{\bfseries}

\begin{table*}[t]
\renewcommand{\arraystretch}{1.2}
\newcommand\setrow[1]{\gdef\rowmac{#1}#1\ignorespaces}
\caption{Results on the LOL test set for different LLIE methods ($\eta$), with and without post-processing. }

\label{tab:lol-results}
\centering
\resizebox{\textwidth}{!}{
\includegraphics{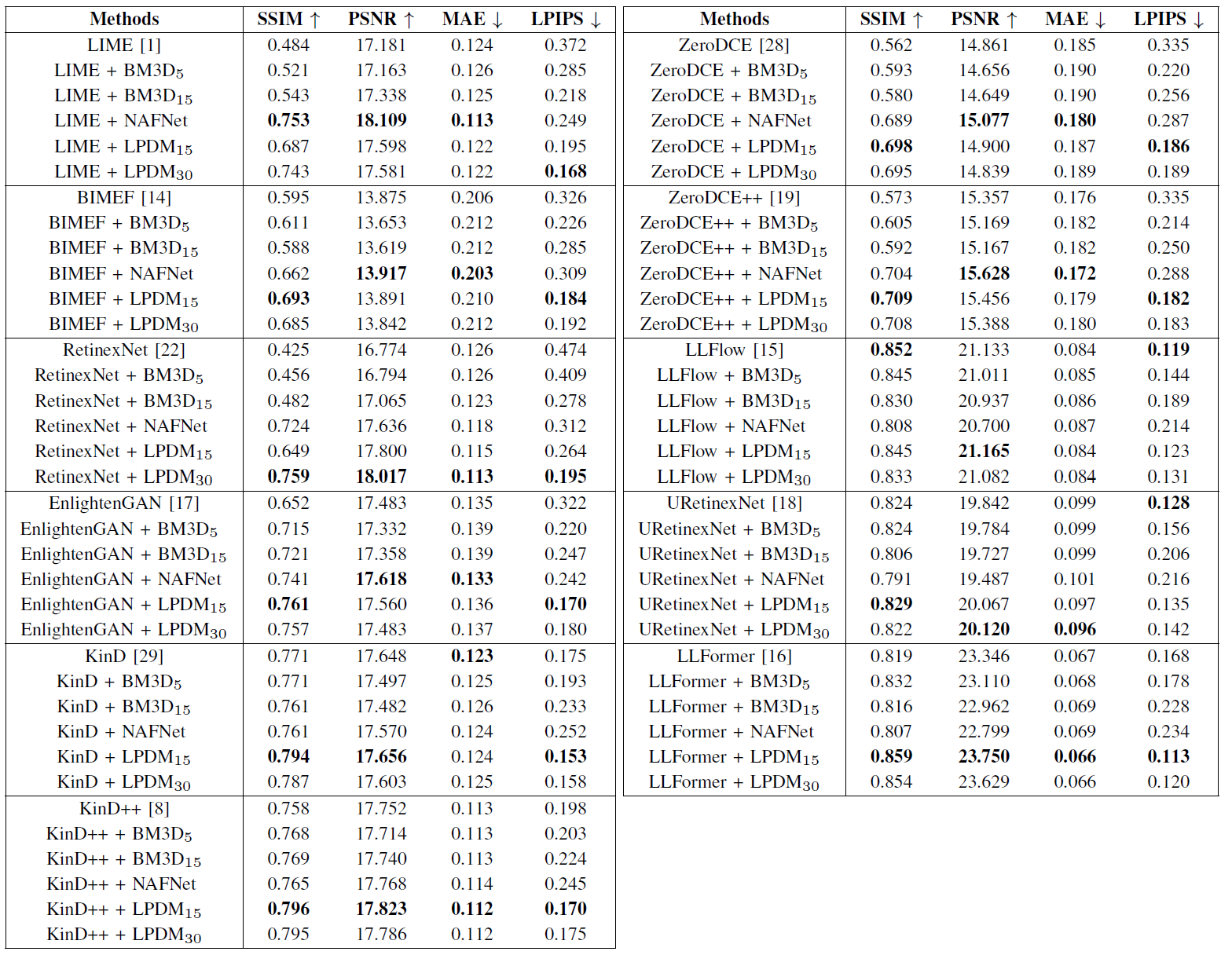}
}
\end{table*}

\begin{figure*}
  \setlength\tabcolsep{2pt}
  \small
  \centering
  \resizebox{\textwidth}{!}{%
  \begin{tabular}{ccccccc}
   LIME \cite{ref:LIME} &
   $\textnormal{BM3D}_5$ &
   $\textnormal{BM3D}_{15}$ &
   NAFNet &
   $\textnormal{LPDM}_{15}$ &
   $\textnormal{LPDM}_{30}$ &
   $\bm{x}_0$ \\
   \includegraphics[width=0.9in]{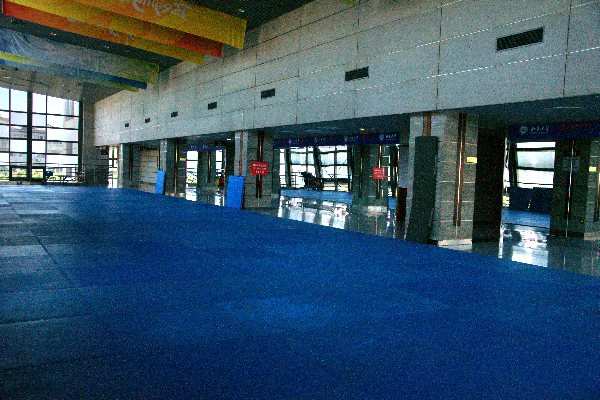} &
   \includegraphics[width=0.9in]{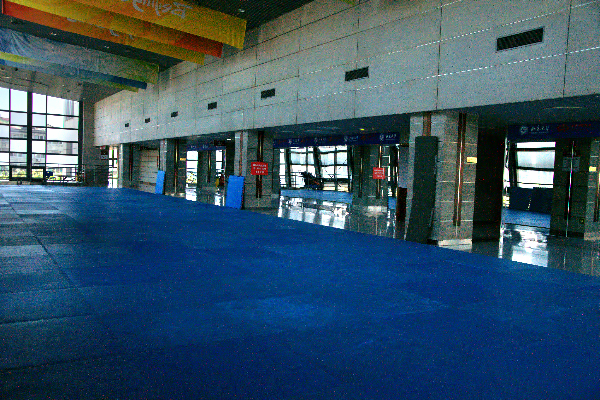} &
   \includegraphics[width=0.9in]{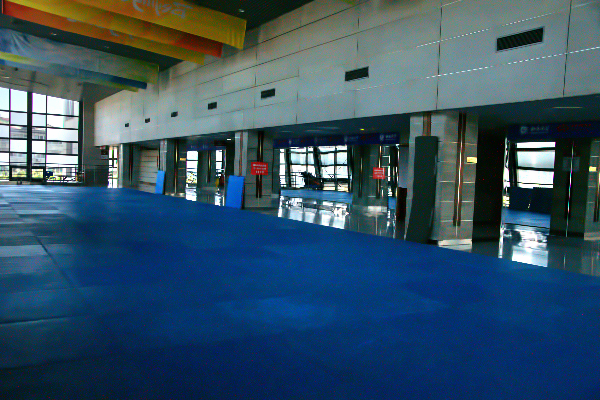} &
   \includegraphics[width=0.9in]{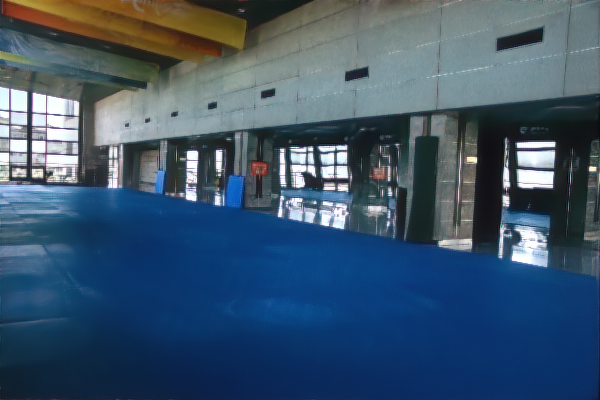} &
   \includegraphics[width=0.9in]{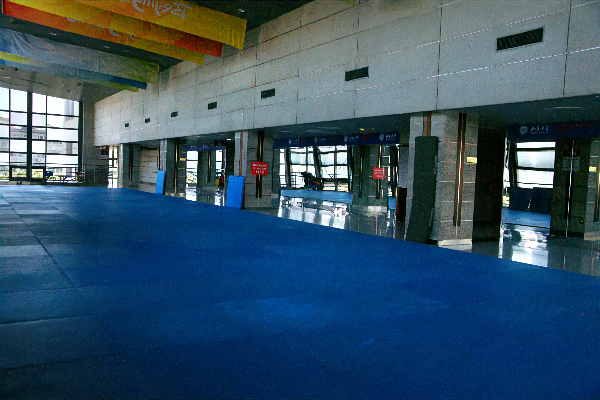} &
   \includegraphics[width=0.9in]{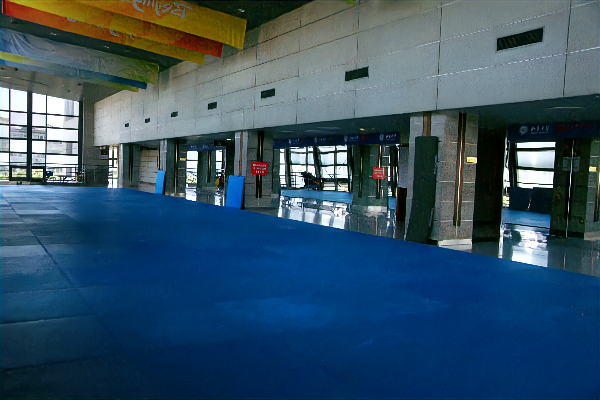} &
   \includegraphics[width=0.9in]{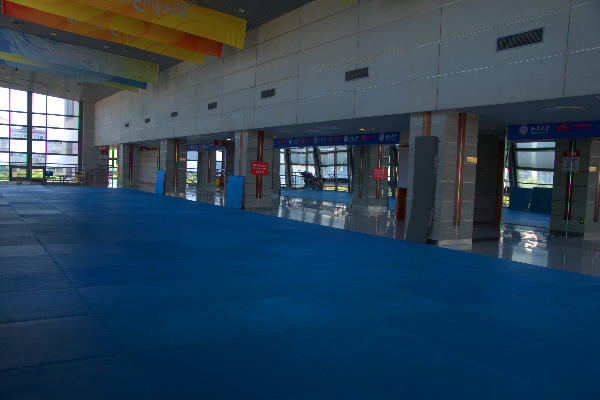} \\
   RetinexNet \cite{ref:lol-dataset} &
   $\textnormal{BM3D}_5$ &
   $\textnormal{BM3D}_{15}$ &
   NAFNet &
   $\textnormal{LPDM}_{15}$ &
   $\textnormal{LPDM}_{30}$ &
   $\bm{x}_0$ \\
   \includegraphics[width=0.9in]{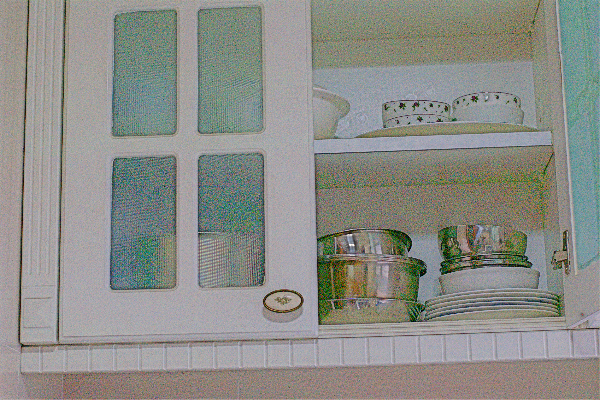} &
   \includegraphics[width=0.9in]{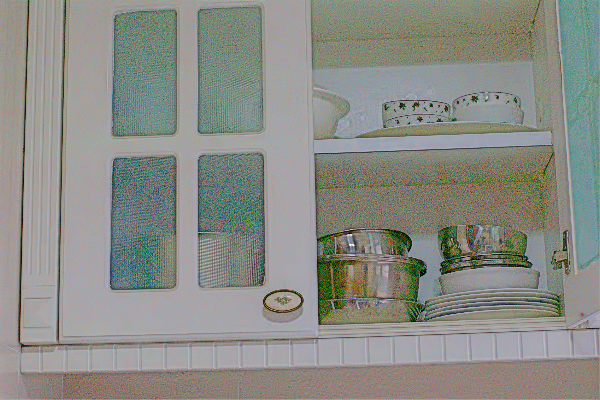} &
   \includegraphics[width=0.9in]{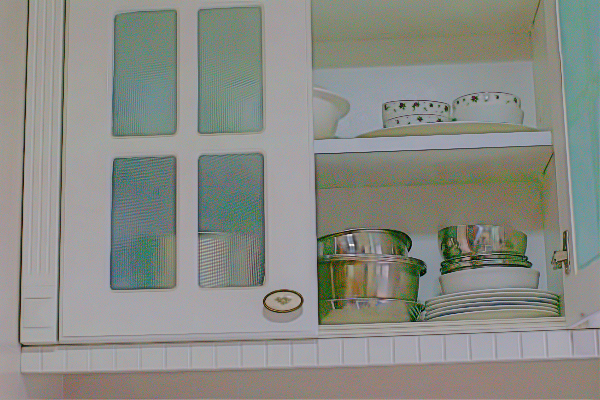} &
   \includegraphics[width=0.9in]{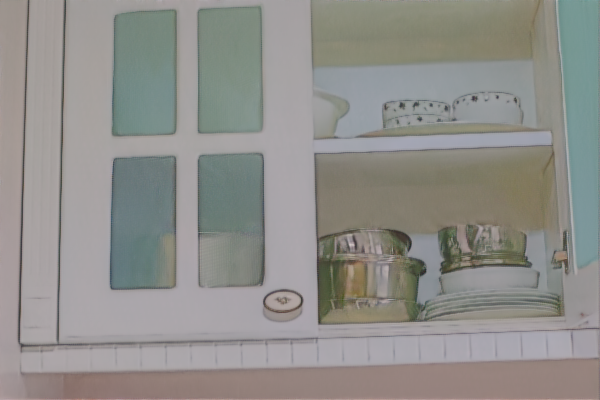} &
   \includegraphics[width=0.9in]{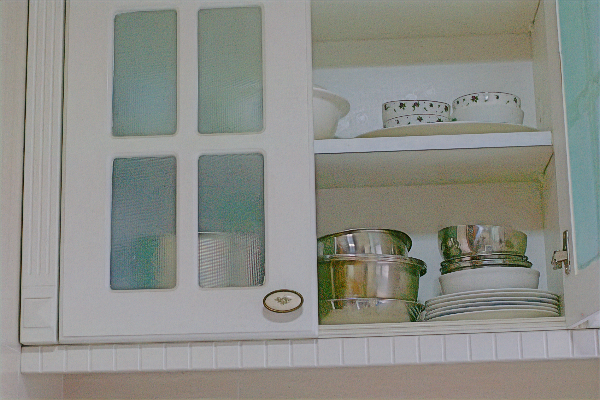} &
   \includegraphics[width=0.9in]{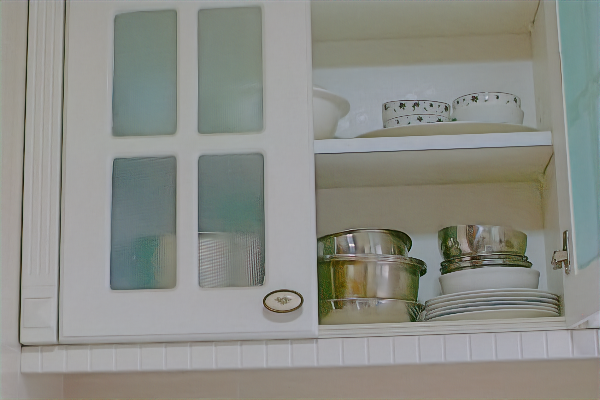} &
   \includegraphics[width=0.9in]{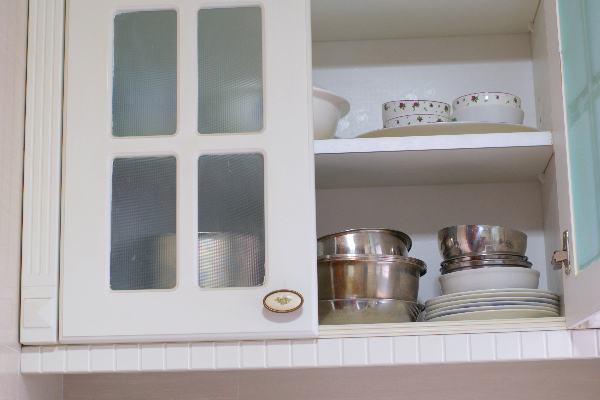} \\
   KinD++ \cite{ref:kind++}&
   $\textnormal{BM3D}_5$ &
   $\textnormal{BM3D}_{15}$ &
   NAFNet  &
   $\textnormal{LPDM}_{15}$ &
   $\textnormal{LPDM}_{30}$ &
   $\bm{x}_0$ \\
   \includegraphics[width=0.9in]{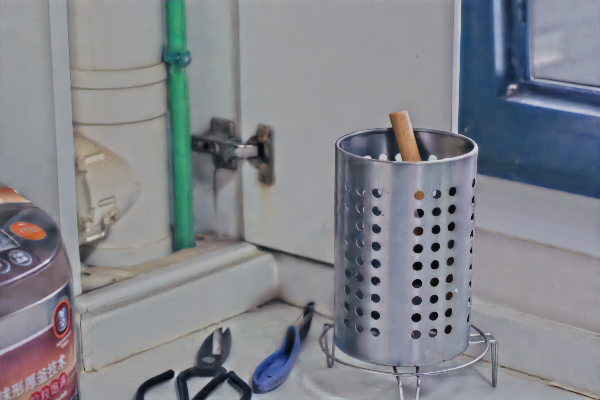} &
   \includegraphics[width=0.9in]{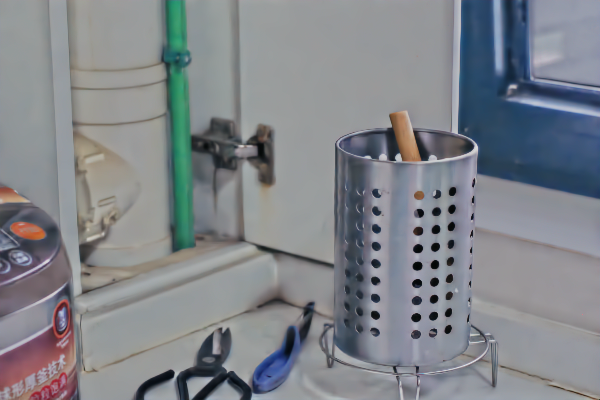} &
   \includegraphics[width=0.9in]{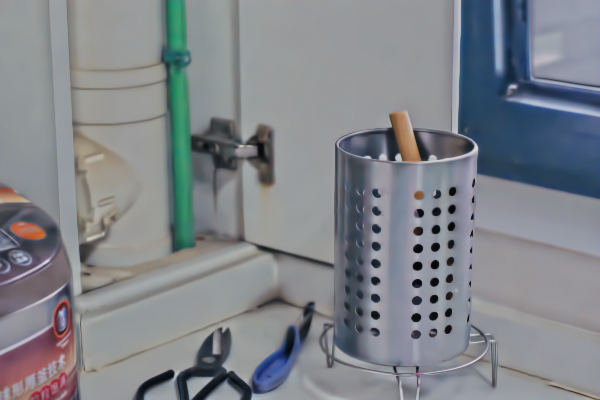} &
   \includegraphics[width=0.9in]{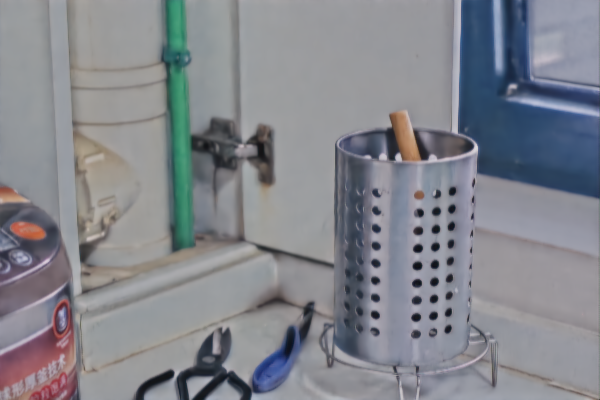} &
   \includegraphics[width=0.9in]{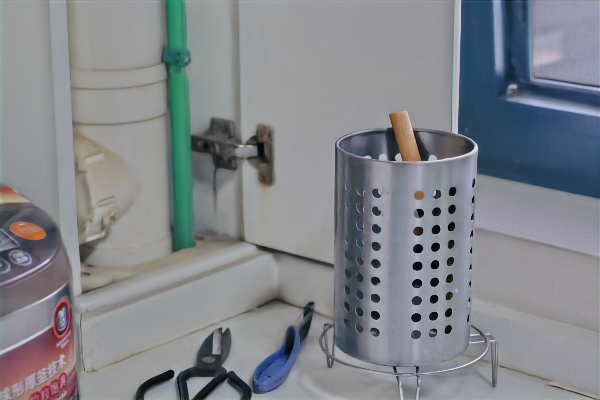} &
   \includegraphics[width=0.9in]{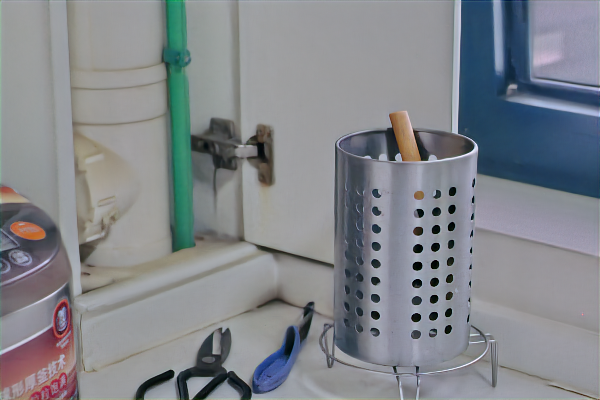} &
   \includegraphics[width=0.9in]{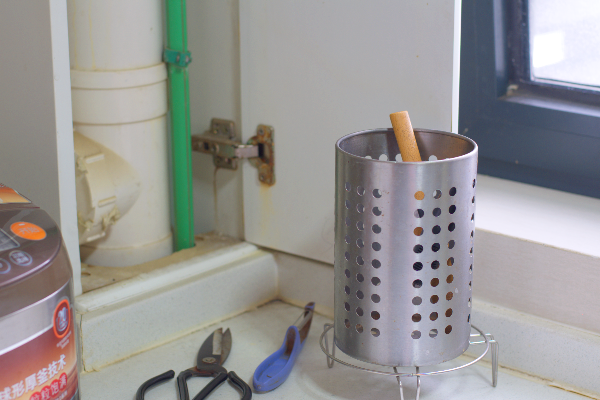} \\
  LLFormer \cite{ref:llformer} &
   $\textnormal{BM3D}_5$ &
   $\textnormal{BM3D}_{15}$ &
   NAFNet  &
   $\textnormal{LPDM}_{15}$ &
   $\textnormal{LPDM}_{30}$ &
   $\bm{x}_0$ \\
   \includegraphics[width=0.9in]{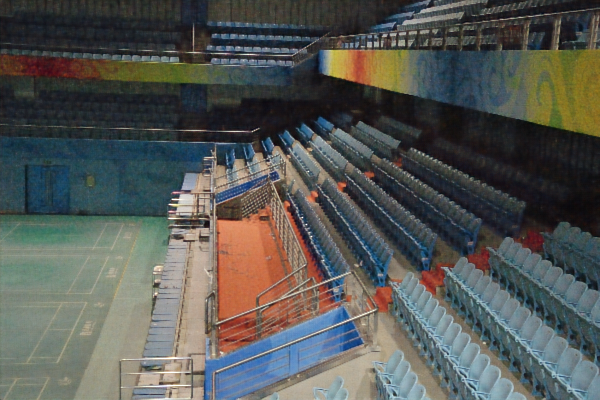} &
   \includegraphics[width=0.9in]{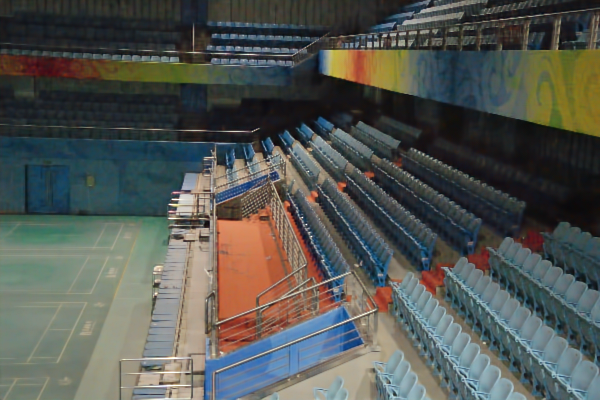} &
   \includegraphics[width=0.9in]{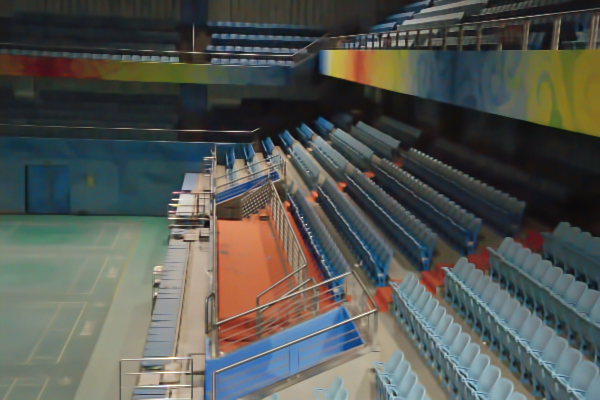} &
   \includegraphics[width=0.9in]{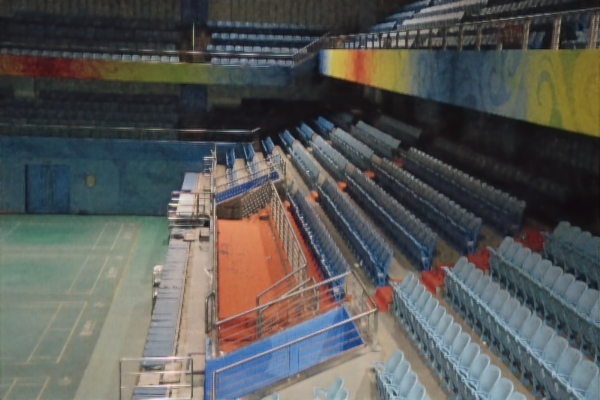} &
   \includegraphics[width=0.9in]{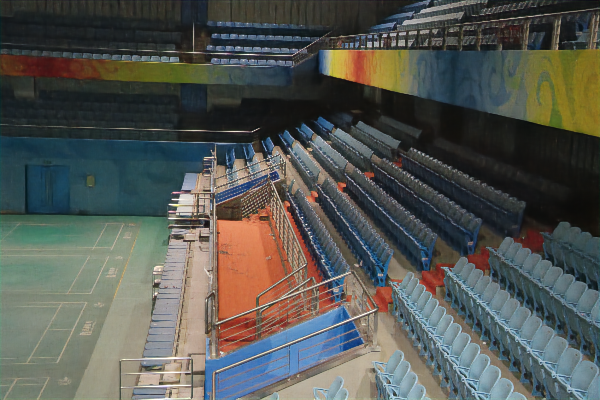} &
   \includegraphics[width=0.9in]{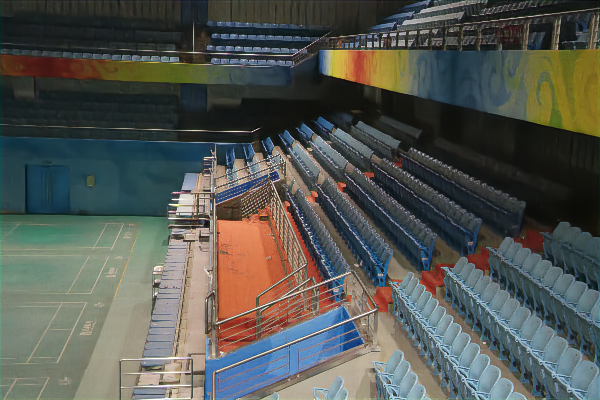} &
   \includegraphics[width=0.9in]{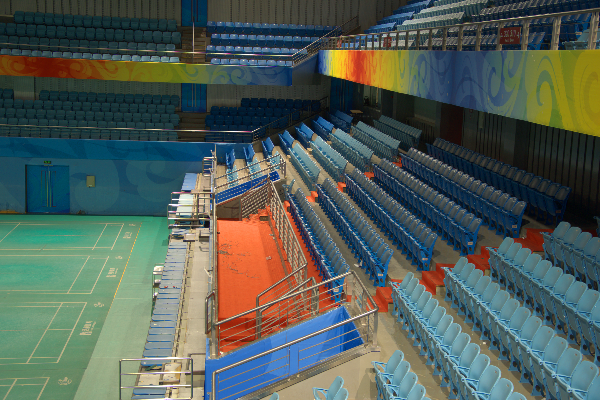} 
  \end{tabular}
  }
  \caption{A qualitative comparison of the BM3D \cite{ref:bm3d} and NAFNet \cite{ref:nafnet-denoiser} post-processing denoising approaches to LPDM on the LOL test set. The first column displays $\bm{\hat{x}}_0^\eta$ for different $\eta$. The final column contains the ground truth label. The remaining columns display the results of different denoising approaches which post-process $\bm{\hat{x}}_0^\eta$.}
  \label{fig:denoiser-compare}
\end{figure*}

\subsection{Comparison with Alternative Denoisers}
\label{sec:experiment:alt-denoisers}
We compare the LPDM denoising performance with the popular BM3D \cite{ref:bm3d} algorithm as well as a state-of-the-art DL denoiser NAFNet \cite{ref:nafnet-denoiser}.  Applying BM3D evenly over the entire image results in bright regions being oversmoothed, since noise is often present at higher levels in dark regions. Therefore, previous LLIE works have adapted BM3D to operate unevenly based on the illumination of the recovered image \cite{ref:lol-dataset,ref:LIME}.

In particular, we test our method against the denoising approach used in LIME \cite{ref:LIME}. LIME is a Retinex-based method where the assumption is that $\bm{\mathrm{L}} = \bm{\mathrm{R}} \circ \bm{\mathrm{T}}$, where $\bm{\mathrm{L}}$ represents the source low-light image, $\bm{\mathrm{R}}$ represents the desired recovery image (reflectance), $\bm{\mathrm{T}}$ represents the illumination and $\circ$ represents element-wise multiplication. LIME estimates the illumination $\bm{\mathrm{T}}$ which is then used to determine $\bm{\mathrm{R}}$ given $\bm{\mathrm{L}}$. In order to avoid oversmoothing bright regions of $\bm{\mathrm{R}}$, the BM3D algorithm is scaled based on the illumination map $\bm{\mathrm{T}}$. Furthermore, the BM3D algorithm is only applied to the Y component of $\bm{\mathrm{R}}$, after $\bm{\mathrm{R}}$ is converted to the YUV color space. The LIME denoising algorithm is as follows \cite{ref:LIME}:

\begin{equation}
  \begin{aligned}
    \bm{\mathrm{R}}_f := \bm{\mathrm{R}} \circ \bm{\mathrm{T}} + \bm{\mathrm{R}}_d \circ (\bm{1} - \bm{\mathrm{T}}),
  \end{aligned}
  \label{eq:denoising:lime}
\end{equation}

\noindent where $\bm{\mathrm{R}}_d$ is the denoised version of $\bm{\mathrm{R}}$, and $\bm{\mathrm{R}}_f$ is the recomposed image after weighting $\bm{\mathrm{R}}$ and $\bm{\mathrm{R}}_d$ by $\bm{\mathrm{T}}$ and $1 - \bm{\mathrm{T}}$ respectively. Note that only the Y channel of $\bm{\mathrm{R}}$ in the YUV color space is denoised, and the result is then converted back to the RGB color space to form $\bm{\mathrm{R}}_d$. We apply LIME to obtain illumination maps $\bm{\mathrm{T}}$ for each low-light image in the LOL test set. We then use \cref{eq:denoising:lime}, and for each $\eta$ we set $\bm{\mathrm{R}} = \bm{\hat{x}}_0^\eta$ for each LOL test example to obtain the denoised result scaled by $\bm{\mathrm{T}}$. For all figures and tables, $\textnormal{BM3D}_{\sigma}$ represents the application of the BM3D algorithm according to \cref{eq:denoising:lime}, where $\sigma$ is the standard deviation parameter of BM3D.
The results are reported in \cref{tab:lol-results} as $\eta + \textnormal{BM3D}_{\sigma}$.

In a further experiment, we apply a state-of-the-art DL denoiser NAFNet \cite{ref:nafnet-denoiser} as a post-processing step after LLIE. The NAFNet model is trained on the smartphone image denoising dataset (SIDD) \cite{ref:sidd} which contains noisy images captured under multiple lighting conditions. Thus, NAFNet is well-suited to denoise enhanced images of a variety of brightness levels. The input to NAFNet is $\bm{\hat{x}}_0^\eta$, and the result is the denoised output which we capture for each $\eta$ over each test image. The results of NAFNet denoising can be found in \cref{tab:lol-results} represented as $\eta + \textnormal{NAFNet}$.

\subsection{Interpretation of Results}
\label{sec:experiment:interp-results}

The full-reference and no-reference metrics are reported in \cref{tab:lol-results} and \cref{tab:unpaired-results}, respectively. Metrics marked in bold indicate that they are the best for a particular method. The LOL test set results in \cref{tab:lol-results} show that our LPDM is able to improve the SSIM of all baseline LLIE methods except LLFlow on the LOL dataset. In all cases, LPDM improves the PSNR compared to each baseline. In many cases, the SSIM is greatly improved by adding the LPDM. Adding the LPDM to LIME, RetinexNet, EnlightenGAN, ZeroDCE, ZeroDCE++ and LLFormer boasts up to a 53.5\%, 78.65\%, 16.8\%, 24.08\%, 23.82\%, 4.92\% SSIM improvement, respectively. LLFormer yields new state-of-the-art color SSIM results on the LOL dataset when LPDM post-processing is added.

LPDM is able to improve the baseline PSNR for all methods, and mostly outperforms the competing denoisers on the PSNR metric. In all cases, LPDM outperforms the alternative denoisers on the perceptual LPIPS metric. In some cases, NAFNet is able to improve PSNR and MAE more than our LPDM; however, upon further inspection, this is as a result of aggressive denoising and thus oversmoothing. \cref{fig:denoiser-compare} displays a comparison of post-processing approaches. NAFNet removes most typical noises, however at the cost of removing detail. For example, consider the first row of \cref{fig:denoiser-compare}, where technically NAFNet yields a higher PSNR for $\bm{\hat{x}}_0^{\textnormal{LIME}}$, but the results are clearly blurred when compared to LPDM which maintains the sharpness of the original image. The behavior of NAFNet manifests more accurately in the LPIPS metric where the LPDM outperforms NAFNet significantly for all LLIE methods. In addition, the BM3D denoiser performs well, however it is unable to deal with color noises and other distortions as robustly as DL methods.

Notably, NAFNet only performs better than LPDM for SSIM on the noise introduced by the LIME method due to the type of noise being similar to most denoising datasets. In contrast, our method models the conditional distribution between low-light and normally-exposed images, and thus the LPDM can handle a variety of different artifacts and color distortions besides typical noise. An example of a distortion which differs from typical Gaussian noise is the distortion introduced by KinD++. Row three of \cref{fig:denoiser-compare} displays how our LPDM increases the sharpness of $\bm{\hat{x}}_0^{\textnormal{KinD++}}$ where the other denoisers yield oversmoothed results. We emphasize this point because different LLIE methods introduce a panoply of different distortions. 

For the majority of methods and datasets in \cref{tab:unpaired-results}, LPDM is able to improve the SPAQ score. The improvement of the NIQE score and BRISQUE score fluctuates depending on the dataset and the method. Therefore, it is vital to analyze the qualitative effects of LPDM on real test data. Several images from a variety of datasets are displayed in \cref{fig:intro}. The general advantage of LPDM is its ability to strike a balance between smoothing and maintaining sharpness. Due to the distribution of the noise output of LPDM being zero-centered, our approach maintains the perceptual quality of the underlying image and avoids oversmoothing when \cref{eq:diffusion:predict-x0-mod} is applied. In some cases, the LPDM is able to improve color quality and sharpness. Upon close inspection, LPDM alters color shades to more accurately represent reality, as seen for BIMEF, KinD++, LLFormer and URetinex-Net examples in \cref{fig:intro}.

In addition to the above conclusions, the choices of $s = 15$ and $s = 30$ may not necessarily be the optimal values for each LLIE method. Therefore, there may be larger improvements for a different choice of $s$ which can be determined empirically. We fix $s$ in order to demonstrate the possibility of using the LPDM as a blind denoiser.

\begin{table*}
  \renewcommand{\arraystretch}{1.2}
  \newcommand\setrow[1]{\gdef\rowmac{#1}#1\ignorespaces}
  \caption{Results on unpaired test sets for different LLIE methods ($\eta$), with and without our proposed LPDM.}
  \label{tab:unpaired-results}
  \centering
  \resizebox{\textwidth}{!}{%
  \includegraphics{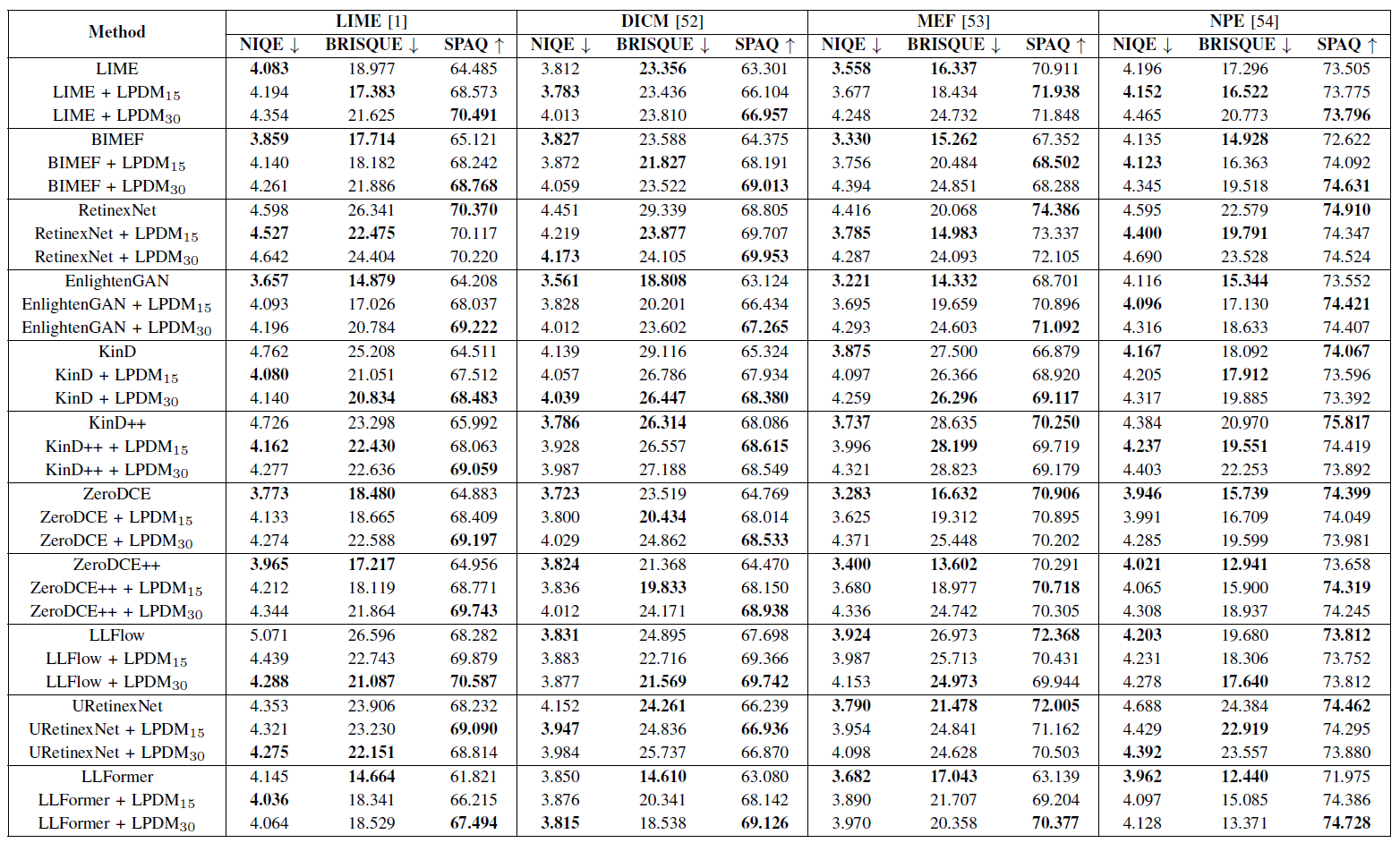}
  }
\end{table*}

\subsection{Ablation Study}
\label{sec:experiment:ablation}
An ablation study is necessary in order to demonstrate that the improvements of LPDM can be attributed specifically to our proposed approach and that the results are not arbitrary. In \cref{sec:experiment:ablation:fixt} we examine the effect of predicting $\bm{\epsilon}$, and in \cref{sec:experiment:ablation:uncond} we compare unconditional diffusion to LPDM. In order to conserve space, tables for the ablation study results are summarized such that the percentage improvement for each metric is calculated for each $\eta$, and the mean and standard deviation percentage improvements are reported. 

\subsubsection{The Effect of Predicting the Noise}
\label{sec:experiment:ablation:fixt}
As seen in \cref{eq:diffusion:ddpm:loss-simple}, DMs are trained to make predictions for $\bm{\epsilon}$. We examine the value of predicting $\bm{\epsilon}$ by changing the model to predict $\bm{x}_0$ directly, and we name this model \textit{direct} LPDM or DLPDM. The DLPDM model directly denoises the input and does not require any further steps such as \cref{eq:diffusion:predict-x0-mod}. The DLPDM is identical to the LPDM described in \cref{sec:experiment:implementation} with two differences: the ground truth target of the model is now $\bm{x}_0$ rather than $\bm{\epsilon}$, and we remove timestep conditioning from the model by setting $t=0$ as input to the model regardless of $\bm{x}_t$. Therefore, the model directly denoises its input with the same number of parameters and without the requirement of specifying $\phi$ at inference time, thus making the model a blind denoiser. In other terms, the model is responsible for detecting the amount of noise present in $\bm{x}_t$ and predicting $\bm{x}_0$ without any additional parameters defined by the user.

\begin{figure}[t]
    \setlength\tabcolsep{2pt}
    \small
    \centering
    \begin{tabular}{cccc}
     Input &
    KinD++ \cite{ref:kind++} &
     DLPDM &
     $\textnormal{LPDM}_{15}$
     \\
     \includegraphics[width=0.8in]{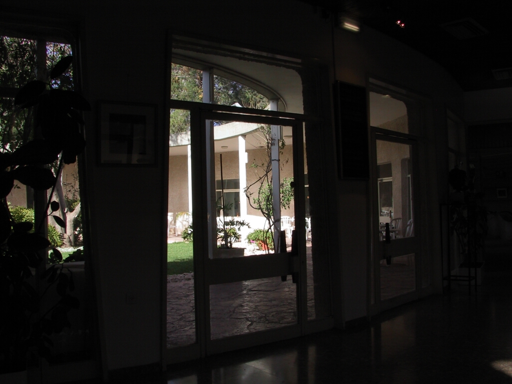} &
     \includegraphics[width=0.8in]{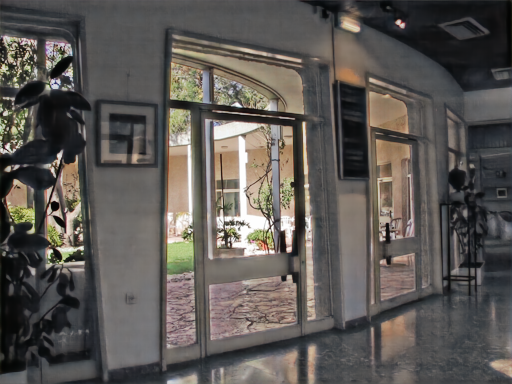} &
     \includegraphics[width=0.8in]{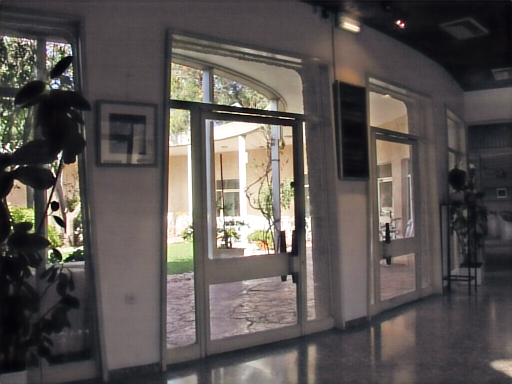} & 
     \includegraphics[width=0.8in]{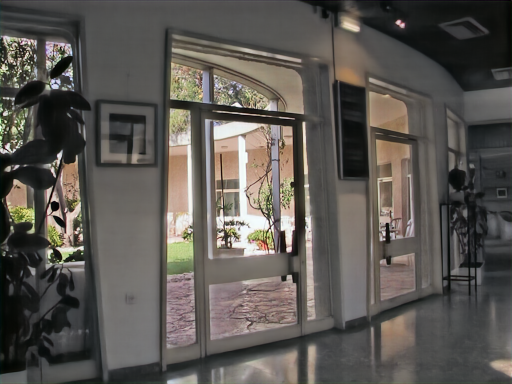} \\ 
     Input &
     RetinexNet \cite{ref:lol-dataset }&
     DLPDM &
     $\textnormal{LPDM}_{15}$
     \\
     \includegraphics[width=0.8in]{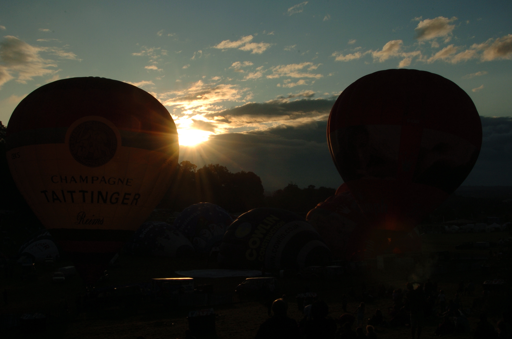} &
     \includegraphics[width=0.8in]{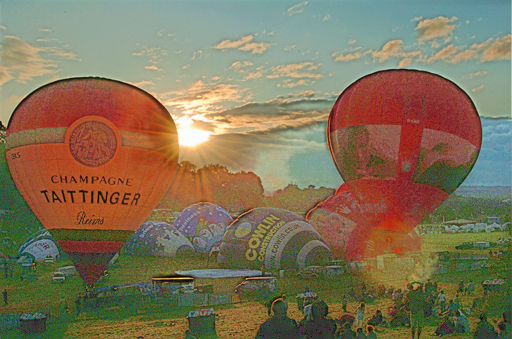} &
     \includegraphics[width=0.8in]{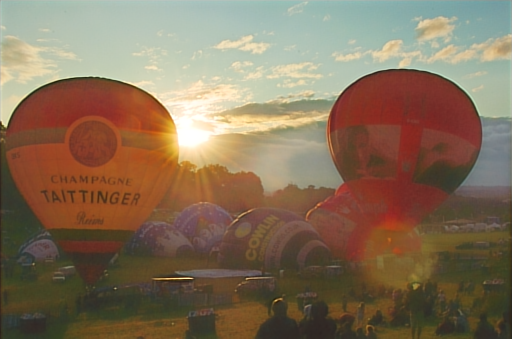} & 
     \includegraphics[width=0.8in]{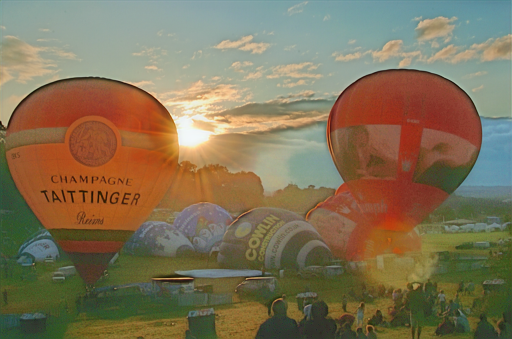}
    \end{tabular}
    \caption{Visual examples of the ablation study directly comparing  predicting $\bm{x}_0$ to predicting $\bm{\epsilon}$ using the DLPDM and LPDM models, respectively.}
    \label{fig:ablation-fixt}
  \end{figure}

The results of the DLPDM experiment are summarized in \cref{tab:ablation-fixt}, which includes the other ablation results from \cref{sec:experiment:ablation:uncond}. Our proposed LPDM approach performs better on SSIM and LPIPS and DLPDM performs better on PSNR and MAE (although the variance of DLPDM is higher). LPDM largely outperforms DLPDM on LPIPS which implies that  LPDM results are more perceptually similar to the ground truth. The results are verified when examining the examples in \cref{fig:ablation-fixt} where LPDM preserves the sharpness of $\bm{\hat{x}_0}^\eta$ and maintains color accuracy.

\begin{table}[t]
  \renewcommand{\arraystretch}{1.2}
  \newcommand\setrow[1]{\gdef\rowmac{#1}#1\ignorespaces}
  \caption{Ablation study comparing LPDM to DLPDM and ULPDM on the LOL test set.}
  
  \label{tab:ablation-fixt}
  \centering
  \resizebox{\linewidth}{!}{
  \begin{tabular}[t]{ 
      |c| 
      c  
      c  
      c  
      c | 
    } 
  \hline
    \B Methods & \multicolumn{1}{c}{\B SSIM (\%) $\uparrow$} & \multicolumn{1}{c}{ \B PSNR (\%) $\uparrow$} & \multicolumn{1}{c}{\B MAE (\%) $\uparrow$} & \multicolumn{1}{c |}{\B LPIPS (\%) $\uparrow$} \\
    \hline
    
    DLPDM &  $19.218 \pm 26.84$ &    $2.53 \pm 3.4$ &  $2.531 \pm 4.93$ &$18.798 \pm 23.76$ \\
    $\textnormal{ULPDM}_{15}$ & $1.937 \pm 28.55$ &$-0.317 \pm 2.42$ &$-0.687 \pm 3.33$ &$-8.43 \pm 39.76$ \\
    $\textnormal{ULPDM}_{30}$ & $-17.566 \pm 28.42$ &  $-2.096 \pm 3.27$ &  $-2.754 \pm 4.95$ &$-43.713 \pm 53.83$ \\
  $\textnormal{LPDM}_{15}$  & $17.138 \pm 17.50$ &  $1.225 \pm 1.79$ &  $0.942 \pm 2.92$ & $29.462 \pm 20.96$ \\
  $\textnormal{LPDM}_{30}$ & $19.928 \pm 25.17$ &  $1.079 \pm 2.26$ &  $0.657 \pm 3.66$ & $28.820 \pm 24.94$ \\
  \hline

    \end{tabular}
  }
  \end{table}

\subsubsection{The Effect of Conditioning}
\label{sec:experiment:ablation:uncond}
We explore the effect of appending $\bm{c}$ to $\bm{x}_t$ as visually depicted in \cref{fig:train-infer}. We name this model \textit{unconditional} LPDM or ULPDM. In many cases, diffusion models may ignore the concatenated conditioning and simply learn how to denoise. Therefore, it is important to explore whether the LPDM requires the use of conditioning to achieve the desired results. ULPDM is an identical model to LPDM from \cref{sec:experiment:implementation}, however, we change the input layer to accept only $\bm{x}_t$ as input, thus changing the number of input channels from six to three. In other terms, we compare conditional diffusion to unconditional diffusion. 

\begin{figure}
    \setlength\tabcolsep{2pt}
    \small
    \centering
    \begin{tabular}{cccc}
     Input &
     LIME \cite{ref:LIME} &
     $\textnormal{ULPDM}_{15}$ &
     $\textnormal{LPDM}_{15}$
     \\
     \includegraphics[width=0.8in]{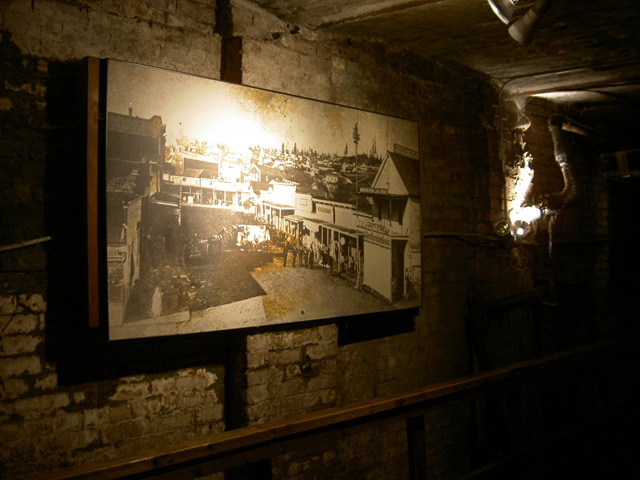} &
     \includegraphics[width=0.8in]{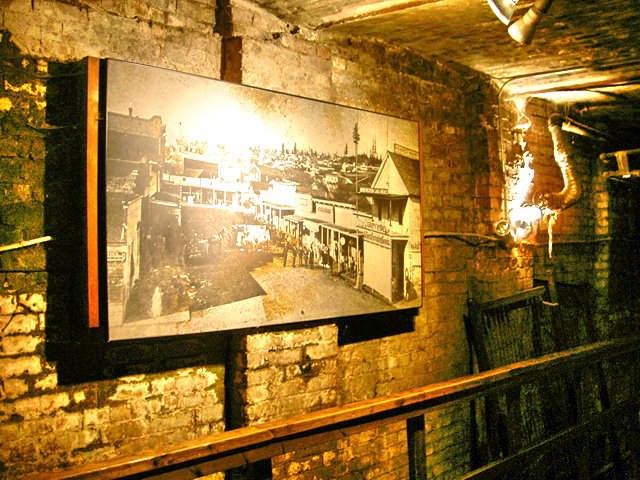} &
     \includegraphics[width=0.8in]{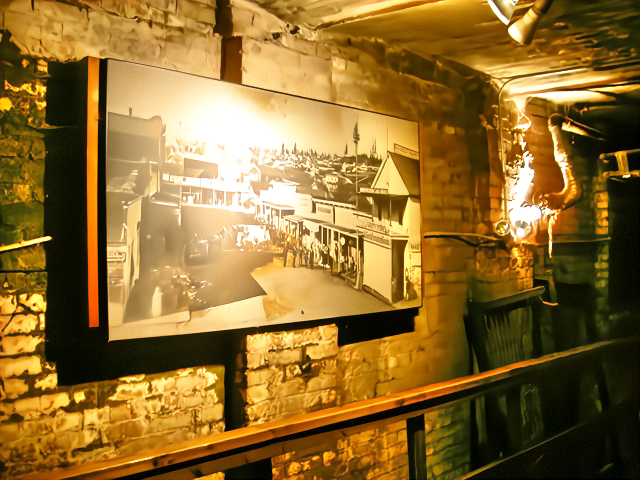} & 
     \includegraphics[width=0.8in]{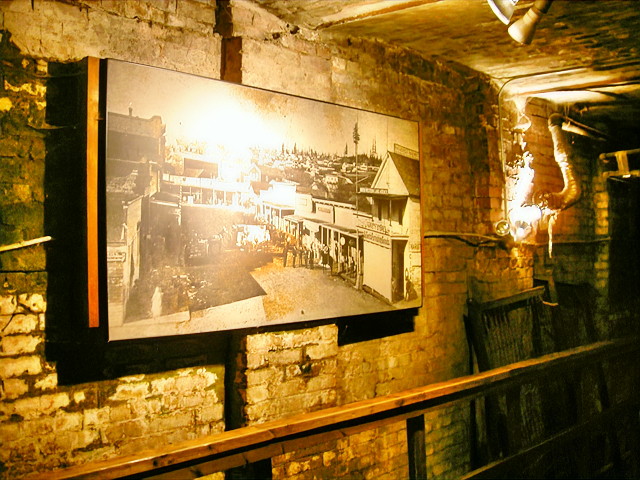} \\ 
     Input &
     LLFlow \cite{ref:llflow} &
     $\textnormal{ULPDM}_{15}$ &
     $\textnormal{LPDM}_{15}$
     \\
     \includegraphics[width=0.8in]{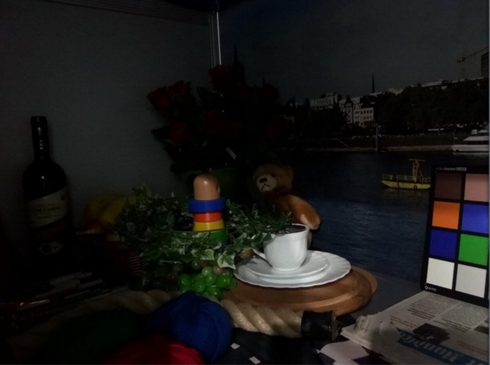} &
     \includegraphics[width=0.8in]{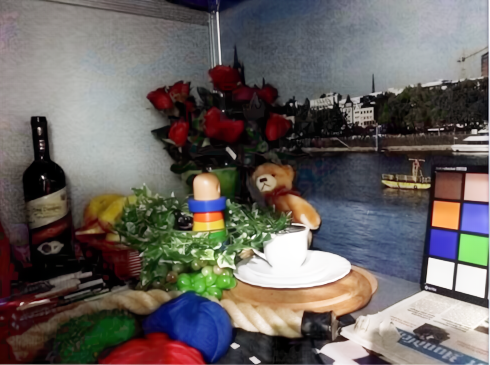} &
     \includegraphics[width=0.8in]{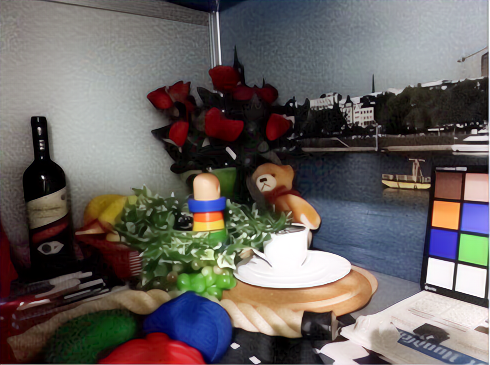} & 
     \includegraphics[width=0.8in]{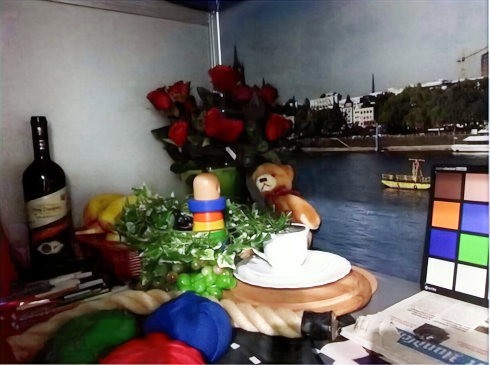}
    \end{tabular}
    \caption{Visual examples of the ablation study comparing unconditional and conditional diffusion using the ULPDM and LPDM models, respectively.
    }
    \label{fig:ablation-uncond}
  \end{figure}

The experimental results in \cref{tab:ablation-fixt} show that LPDM significantly outperforms ULPDM across all metrics. Therefore, we conclude that conditioning is necessary in order for the LPDM to detect the wide variety of artifacts that can be present in $\bm{\hat{x}}_0^\eta$. We provide visual results in \cref{fig:ablation-uncond} which verify our conclusion: ULPDM is able to remove noise, however results are oversmoothed and thus detail is lost due to lack of conditioning.

\section{Conclusion}
\label{sec:conclusion}
In this paper, we present a framework for post-processing images which have undergone low-light image enhancement. The enhancement of low-light images often reveals a variety of degradations which are hidden in the dark, and thus a need for post-processing is introduced. Furthermore, each low-light enhancement technique can possibly introduce a different form of degradation into its result. We propose using a conditional diffusion model in order to model the distribution between under-exposed and normally-exposed images. Further, we introduce a method of applying the diffusion model as a post-processing technique. Our approach uses the diffusion model to estimate the amount of noise present in an enhanced image in one pass through the model, which can simply be subtracted from the enhanced image to further enhance the image. Moreover, we demonstrate that our approach outperforms competing post-processing denoisers, and we demonstrate its versatility on a variety of low-light datasets with different state-of-the-art low-light image enhancement backbones. In contrast to existing denoisers, we find that our approach is able to improve perceptual quality, while removing noise and other distortions. In future work, our approach could potentially be applied to other image restoration domains.


%

\ifCLASSOPTIONcompsoc
  \section*{Acknowledgments}
\else
  \section*{Acknowledgment}
\fi

This study was supported by the National Research Foundation (NRF), South Africa, Thuthuka Grant Number 13819413. The authors acknowledge the Centre for High Performance Computing (CHPC), South Africa, for providing computational resources to this research project.

\ifCLASSOPTIONcaptionsoff
  \newpage
\fi



\bibliographystyle{IEEEtran}
\bibliography{IEEEabrv, references}
\end{document}